\documentclass[aps,prb,floatfix,epsfig,twocolumn,showpacs,preprintnumbers]{revtex4}
\UseRawInputEncoding
\usepackage{amsfonts}
\usepackage{amssymb}
\usepackage{graphicx}
\usepackage{amsmath}
\usepackage{tablefootnote}
\usepackage{setspace}
\begin{document}

\title{Elimination of the linearization error in APW/LAPW basis set: Dirac-Kohn-Sham equations}

\author{Andrey L. Kutepov\footnote{e-mail: akutepov@bnl.gov}}

\affiliation{Condensed Matter Physics and Materials Science Department, Brookhaven National Laboratory, Upton, NY 11973}

\begin{abstract}
A detailed account of the implementation of equations of the Relativistic Density Functional Theory (RDFT) using basis sets of APW/LAPW type with flexible extensions provided by local orbitals is given. Earlier discoveries of the importance of the High Derivative Local Orbital (HDLO) extension of APW/LAPW basis set for enhancing the accuracy of DFT calculations are confirmed using fully relativistic approach and $\alpha$-U as an example. High Energy Local Orbitals (HELO's), however indispensable for GW calculations, are considerably less efficient in enhancing the accuracy of DFT applications. It is shown, that a simplified approach to the relativistic effects, namely, considering them only inside the muffin-tin (MT) spheres, produces basically identical results (as compared to fully relativistic approach) for the electronic free energy of the five materials considered in this work. By comparing the effect of the simplified approach on the electronic free energy with its effect on the electronic kinetic energy we conclude that the insensitivity of the free energy to the way we describe the relativistic effects in the interstitial region is related to the variational property of this quantity.
\end{abstract}




\maketitle



\section{Introduction}
\label{intr}

The importance of correct treatment of relativistic effects in solids was well known already at the beginning of the era of computer codes creation. In a general context of the Density Functional Theory (DFT) \cite{pr_136_B864}, its equations were generalized to fully relativistic form by Rajagopal, Callaway, Vosko and Ramana \cite{prb_7_1912,jpc_11_L943,jpc_12_2977,jpc_12_L845}. In the context of specific basis set formulations, an introduction of a new methodology for calculation of electronic band structure (usually in non-relativistic form) was quickly followed by its generalization to fully relativistic form based on Dirac equation. Loucks \cite{pr_139_A1333} generalized non-linear Augmented Plane Wave (APW) method originally proposed by Slater \cite{pr_51_846}. Another popular method with a long history, KKR, proposed by Korringa, Kohn, and Rostoker \cite{phys_13_392,pr_94_1111} was generalized by Huhne et al. \cite{prb_58_10236} to fully relativistic and full potential form. Introduction of the linearized augmented plane wave (LAPW) and linearized muffin-tin orbitals (LMTO) band structure methods by O.K. Andersen \cite{prb_12_3060} was followed by generalizations of LAPW (Takeda \cite{jpf_9_815}, MacDonald et al. \cite{jpc_13_2675}) and of LMTO (Ebert et al. \cite{jap_63_3052}).

Relativistic formulation of the band structure approach makes it, however, a bit more complicated as the corresponding equations are considerably more involved than their non-relativistic counterparts. This fact makes it more difficult to include effects beyond DFT, such as, for instance, Hedin's GW approximation \cite{pr_139_A796} if one works in fully relativistic framework. Also, the size of the basis set is normally doubled in the relativistic case which means nearly tenfold increase of the computer time needed to accomplish the calculation. These circumstances lead the development of the band structure methods in the last 20-30 years mostly along the path of increasing efficiency and new effects were added mostly at non-relativistic level. Thus, in terms of proper accounting of the relativistic effects there was a certain sacrifice of accuracy as many codes accepted a "standard" approach which is scalar relativistic (SR) approximation (total neglect of spin-orbit coupling (SOC)) with optional treatment of SOC as a perturbation (abbreviation SR+SOC). For instance, most popular and most cited codes such as WIEN2k \cite{wien2k} and VASP \cite{vasp} were developed along this path. Certainly, for many materials where relativistic effects are weak the above simplification is well justified. Its validity, however, is doubtful for materials which have atoms with heavy nucleus, such as actinides. For instance, total inadequacy of SR+SOC when addressing the equilibrium volume in actinides was discovered some time ago \cite{prb_63_035103}. The issue revealed itself as strong dependence of the calculated total energy (and particularly its volume dependence) on the selected muffin-tin radius, making it practically impossible to predict unique equilibrium volume in actinides using SR+SOC approach to relativistic effects. A simple solution of the problem which seems to work well for actinides is to discard SOC for the $p$ states \cite{as_9_5020}. More involved solution is to include the Dirac $6p_{1/2}$ orbitals to the basis set of SR orbitals \cite{prb_64_153102}. This remedy (we will use abbreviation $SR+SOC+p_{1/2}$ for it), though quite useful (as subsequent applications have shown), brings certain in-homogeneity in the basis set making it difficult for further developments (such as GW formulation). Also its accuracy as compared to full Dirac equation implementation is not quite clear, because the effect of SR approximation for the rest (excluding $6p_{1/2}$ orbitals) of the basis set has not been yet thoroughly investigated. For instance, recent research \cite{prm_1_033803} was using $SR+SOC+p_{1/2}$ as a standard to test other approximations for SOC (such as non-self-consistent treatment). Another recent work \cite{prb_101_085114} introduces Dirac equation based implementation of the LMTO method, but compares it only to SR+SOC results, avoiding comparison to $SR+SOC+p_{1/2}$ results.

Thus, the importance of development (and updating of the already existing) computer codes based on Dirac equation should be obvious. Their role is not confined to the DFT realm. More advanced theories (such as GW approximation) or DFT+DMFT \cite{rmp_68_13} when applied to materials with strong relativistic effects should include relativistic effects on equal footing with the effects of electron correlation.

This work was mostly inspired by recent developments aimed at the elimination of the linearization error in approaches based on the LAPW basis set. Local orbitals extensions of LAPW designed to describe semicore and high energy states \cite{prb_43_6388} (we will call them as High Energy Local Orbitals - HELO) have been known for more than 25 years and was included in older versions of the FlapwMBPT code \cite{jcm_15_2607}. Important recent advances such as APW+lo extension \cite{prb_64_195134} and High Derivative Local Orbitals (HDLO) extension \cite{prb_74_045104,cpc_184_2670,cpc_220_230} brought, however, considerable enhancements in the accuracy of the LAPW/APW schemes basically eliminating the linearization error. They, to the best of the author's knowledge, have not been formulated for the Dirac equation yet.

This work pursues two principal goals. First, it aims at filling the gap in fully relativistic generalizations of modern approaches to the electronic structure of solids. Namely, the recent advances, APW+lo and HDLO, are formulated for the Dirac equation in a flexible combination where APW, APW+lo, and LAPW can be supplemented with HDLO or HELO in different ways depending on L-channel and atom. This formulation is based on the older implementation with the LAPW+HELO basis set \cite{jcm_15_2607}. Present work particularly stresses the importance of the Schlosser-Marcus (SM) variational principle \cite{pr_131_2529,pr_139_A1333} which becomes especially useful when one goes from the LAPW+HELO basis set (which is only slightly discontinued at the muffin-tin boundaries via the finite number of spherical harmonics in the expansion inside MT) towards APW+lo (every term in spherical harmonics expansion has discontinuities). Whereas older work \cite{jcm_15_2607} was also based on this principle, it was not critical and, correspondingly, was not specifically addressed. This goal includes thorough testing of the implementation and its accuracy using $\alpha$-uranium as an example.

The second goal is to provide evidence of usefulness of an approximate treatment of the full Dirac equation. This approximate form (we will use the term SRA for it, where S stands for Simplified) consists in neglecting the relativistic effects in the interstitial region whereas retaining the full Dirac treatment inside MT. The full Dirac approach will be abbreviated as FRA. SRA retains more relativistic effects as compared to $SR+SOC+p_{1/2}$ which not only accepts non-relativistic description in the interstitial region but also uses an additional approximations inside MT. The SRA approach was in fact used in our formulation of fully self-consistent GW method \cite{prb_85_155129} and the GW+Vertex scheme \cite{prb_94_155101} has also been generalized to SRA level \cite{unpubl}, but its comparison with the full Dirac approach was never (even at DFT level) published. The second goal includes comparison with the results \cite{prm_1_033803} obtained with $SR+SOC$ and $SR+SOC+p_{1/2}$ for a few compounds and, by doing it, to provide more information on the quality of $SR+SOC+p_{1/2}$ approximation. Finally we will compare calculated magnetic anisotropy energies (MAE) for FePt with results available in the literature (on both levels - full Dirac and simplified) to make conclusions on the validity of the simplified form in the applications to MAE.

\section{Methodology}
\label{GW}

\subsection{Relativistic Spin-Polarized Density Functional Theory} \label{RDFT}

An approximation for a joint description of relativistic and magnetic effects within Relativistic Density Functional Theory (RDFT), convenient for realistic applications, was developed in works by Rajagopal, Callaway, Vosko and Ramana \cite{prb_7_1912,jpc_11_L943,jpc_12_2977,jpc_12_L845}. Essential ingredient of this formalism consists in the so called no-pair approximation, i.e. in the neglect of all effects related to the existence of negative states in the relativistic theory. Another important step is splitting of the total four-current (by use of the Gordon decomposition) into the paramagnetic and the spin components and the neglect of the paramagnetic part. Principal equations of this theory are briefly capitalized in this section for completeness. As we will consider the electronic finite temperature, it is natural to begin by writing down the expression for electronic free energy. Electronic free energy of a solid with electronic density
$n(\mathbf{r})$ and magnetization density $\mathbf{m}(\mathbf{r})$ can be written as the following:
\begin{widetext}
\begin{align}
\label{etot} F[n,\mathbf{m}] &= -T\sum_{\mathbf{k}\lambda}\ln (1+e^{-(\epsilon^{\mathbf{k}}_{\lambda}-\mu)/T})+\mu N -\int_{\Omega_{0}} \text{d}
\mathbf{r} [n(\mathbf{r})V_{eff}(\mathbf{r})+\mathbf{m}(\mathbf{r})\cdot
\mathbf{B}_{eff}(\mathbf{r})]+E_{nn}\nonumber\\&+\int_{\Omega_{0}} \text{d}
\mathbf{r} [n(\mathbf{r})V_{ext}(\mathbf{r})+\mathbf{m}(\mathbf{r})\cdot
\mathbf{B}_{ext}(\mathbf{r})] + \int_{\Omega_{0}} \text{d} \mathbf{r}
\int_{\Omega} \text{d} \mathbf{r'} \frac{n(\mathbf{r})n(\mathbf{r'})}{|\mathbf{r}-\mathbf{r'}|}+E_{xc}[n,\mathbf{m}],
\end{align}
\end{widetext}
where $T$ stands for the temperature, sum runs over the Brillouin zone points $\mathbf{k}$ and band indexes $\lambda$, $\epsilon^{\mathbf{k}}_{\lambda}$ is the band energy, $\mu$ is the chemical potential, and $N$ is the total number of electrons in the unit cell. $\Omega_{0}$ is the volume of the primitive unit cell and $\Omega$ is the volume of the whole solid. Atomic units are assumed with Rydbergs being the units of energy. 
Effective scalar potential $V_{eff}(\mathbf{r})$ is a sum of the external scalar field and induced fields (electrostatic and exchange-correlation):
\begin{equation} \label{veff}
V_{eff}(\mathbf{r})=V_{ext}(\mathbf{r})+2 \int_{\Omega} \text{d} \mathbf{r'}
\frac{n(\mathbf{r'})}{|\mathbf{r} - \mathbf{r'}|} + \frac{\delta
E_{xc} [n(\mathbf{r}),\mathbf{m}(\mathbf{r})]}{\delta n(\mathbf{r})},
\end{equation}
whereas the effective magnetic field $\mathbf{B}_{eff}(\mathbf{r})$ represents a sum of external and induced magnetic fields:
\begin{equation} \label{beff}
\mathbf{B}_{eff}(\mathbf{r})=\mathbf{B}_{ext}(\mathbf{r})+ \frac{\delta E_{xc}
[n(\mathbf{r}),\mathbf{m}(\mathbf{r})]}{\delta \mathbf{m}(\mathbf{r})}.
\end{equation}

$E_{nn}$ in (\ref{etot}) is nucleus-nucleus coulomb interaction energy and $E_{xc}$ stands for the exchange-correlation energy which is a functional of $n(\mathbf{r})$ and $\mathbf{m}(\mathbf{r})$.
One-electron energies $\epsilon^{\mathbf{k}}_{\lambda}$ are the eigen values of the following equations (Dirac-Kohn-Sham equations):
\begin{equation} \label{ksh}
\left(\hat{K}+V_{eff}(\mathbf{r})+ \beta
\widetilde{\boldsymbol{\sigma}} \cdot \mathbf{B}_{eff}(\mathbf{r})\right) \Psi_{\lambda}^{\mathbf{k}}(\mathbf{r})=\epsilon^{\mathbf{k}}_{\lambda}
\Psi_{\lambda}^{\mathbf{k}}(\mathbf{r}),
\end{equation}
with the Dirac form of the kinetic energy operator $\hat{K}$ (electron rest energy has been subtracted),
\begin{equation} \label{hkin}
\hat{K}=c \boldsymbol{\alpha} \cdot \mathbf{p} +(\beta-I)
\frac{c^2}{2}.
\end{equation}
$\Psi_{\lambda}^{\mathbf{k}}(\mathbf{r})$ stands for Bloch periodic band function, $c$ is for light velocity ($c=274.074$ in our unit system), $\mathbf{p}$ - momentum operator ($\equiv -i \nabla$),
$\boldsymbol{\alpha}$, $\beta$ are Dirac matrices in standard representation, $I$ is unit $4 \times 4$ matrix.
In (\ref{ksh}), $\widetilde{\boldsymbol{\sigma}}$ are $4 \times 4$ matrices, combined from the Pauli matrices $\boldsymbol{\sigma}$:
\begin{equation}
\label{Pauli}
\widetilde{\boldsymbol{\sigma}}= \left(
\begin{array}{cc}
\boldsymbol{\sigma} & 0 \\
0 & \boldsymbol{\sigma}
\end{array} \right).
\end{equation}

With the electron energies and band state functions available, one can evaluate electronic and magnetization densities as the following
\begin{equation} \label{dens}
n(\mathbf{r})=\sum_{\mathbf{k}\lambda}f^{\mathbf{k}}_{\lambda}
\Psi_{\lambda}^{^{\dag}\mathbf{k}}(\mathbf{r}) \Psi_{\lambda}^{\mathbf{k}}(\mathbf{r}),
\end{equation}
and
\begin{equation} \label{magn}
\mathbf{m}(\mathbf{r})=\sum_{\mathbf{k}\lambda}f^{\mathbf{k}}_{\lambda}
\Psi_{\lambda}^{^{\dag}\mathbf{k}}(\mathbf{r}) \beta
\widetilde{\boldsymbol{\sigma}} \Psi_{\lambda}^{\mathbf{k}}(\mathbf{r}),
\end{equation}
with $f^{\mathbf{k}}_{\lambda}$ being the Fermi-Dirac distribution function ($f^{\mathbf{k}}_{\lambda}=\frac{1}{1+e^{(\epsilon^{\mathbf{k}}_{\lambda}-\mu)/T}}$).

In the expression (\ref{etot}), there are three divergent terms - electrostatic electron-electron, electron-nucleus, and nucleus-nucleus interactions. The divergencies, however, mutually cancel providing us with suitable for evaluation expression:
\begin{eqnarray}\label{etot1}
E_{nn}&+\int_{\Omega_{0}} \text{d}
\mathbf{r} n(\mathbf{r})V_{ext}(\mathbf{r})+ \int_{\Omega_{0}} \text{d} \mathbf{r}
\int_{\Omega} \text{d} \mathbf{r'} \frac{n(\mathbf{r})n(\mathbf{r'})}{|\mathbf{r}-\mathbf{r'}|} \nonumber\\&= \frac{1}{2}\int_{\Omega_{0}} \text{d} \mathbf{r}
n(\mathbf{r})V_{C}(\mathbf{r})- \frac{1}{2} \sum_{t}
Z_{t}V'_{C}(\mathbf{t}),
\end{eqnarray}
where vectors $\mathbf{t}$ represent atomic coordinates, $Z_{t}$ - nucleus charge of atom $t$,
$V_{C}(\mathbf{r})$ is total coulomb potential, and
$V'_{C}(\mathbf{t})$ is coulomb potential at the center of atom $t$ excluding nucleus charge contribution of that atom.

The self-consistent solution of the above equations begins with providing an initial distribution of the electronic density. Normally it is a sum of overlapping atomic densities. Initial magnetization is taken to be zero but in order to obtain possible magnetic solution an initial (acting only at the first iteration) external magnetic field is provided. The effective potential is evaluated according to (\ref{veff}). Then equations
(\ref{ksh}) are solved. Having found the band energies and corresponding Bloch functions, one searches for the chemical potential (to be obtained from electro-neutrality condition) and evaluates $n(\mathbf{r})$ and
$\mathbf{m}(\mathbf{r})$ according to (\ref{dens}) and (\ref{magn}). After that (if the self-consistency is not yet attained) one evaluates $V_{eff}(\mathbf{r})$ and $\mathbf{B}_{eff}(\mathbf{r})$ according to (\ref{veff}) and
(\ref{beff}), which closes the iteration loop.

The equations presented in this section serve as a basis for the present work. A detailed account of the implementation specific to a chosen in the code FlapwMBPT basis set is provided in the following sections.

\subsection{Representation of $n(\mathbf{r}), V(\mathbf{r}),
\mathbf{m(r)}, \mathbf{B(r)}$} \label{represent}

In the LAPW family of the basis sets, the space is divided into non-overlapping muffin-tin (MT) spheres of volume $\Omega_{t}$ and radius $S_{t}$, and the interstitial region with volume $\Omega_{I}$. In accordance with dual representation for Bloch functions (i.e. expansion in spherical harmonics inside the MT spheres and expansion in plane waves in the interstitial region) one uses similar representation for other quantities: charge density $n(\mathbf{r})$, effective scalar potential $V_{eff}(\mathbf{r})$, magnetization density $\mathbf{m(r)}$ and magnetic field
$\mathbf{B(r)}$. For scalar quantities one has, accordingly,
\begin{equation} \label{V_rep}
V(\mathbf{r})=
\begin{cases}
\sum_{\mathbf{G}}V_{\mathbf{G}}e^{i\mathbf{G}\cdot \mathbf{r}} &
\mathbf{r}\in I
\\
\sum_{lm}V_{lm}(r)\overline{Y}_{lm}(\theta, \varphi) &\mathbf{r}\in MT
\end{cases},
\end{equation}
and for the vector quantities similar representation holds for all components of the vector:
\begin{equation} \label{B_rep}
\mathbf{B}(\mathbf{r})=
\begin{cases}
\sum_{\mathbf{G}}\mathbf{B}_{\mathbf{G}}e^{i\mathbf{G}\cdot
\mathbf{r}} & \mathbf{r}\in I
\\
\sum_{lm}\mathbf{B}_{lm}(r)\overline{Y}_{lm}(\theta, \varphi) &\mathbf{r}\in
MT,
\end{cases}
\end{equation}
where we use over-line symbols for real-valued spherical harmonics in order to distinguish them from the complex-valued spherical harmonics.

\subsection{Evaluation of $V_{xc}(\mathbf{r})$ and $\mathbf{B(r)}$}
\label{V_xc_B}

Quantities $V_{xc}(\mathbf{r})$ and $\mathbf{B_{eff}(r)}$ are evaluated in real space. Inside the MT spheres, for each value of the radial variable (distance from the nucleus), the density and the magnetization (also the space derivatives in case if we use Generalized Gradient Approximation - GGA) are evaluated on a grid of two angular variables. Then $V_{xc}(\mathbf{r})$ and $\mathbf{B_{eff}(r)}$ are evaluated on the same grid and their representation in terms of spherical harmonics is found by numerical angular integration. In the interstitial region, one uses similar procedure. The only difference is that 3D regular grid in the unit cell is used and the transformation between the real space and the recirocal space is accomplished with the Fast Fourier Transforms (FFT).

\subsection{Basis functions in fully relativistic implementation of the FlapwMBPT code}\label{basis}

A few different basis functions are generally used in FlapwMBPT. They can be divided in two types - augmented waves and local functions. In order to distinguish between them we will use generic symbol $\Pi$ for augmented waves and symbol $\Lambda$ for local functions. Below these two types of the basis functions are discussed in detail. We will also introduce a simplified approach for the relativistic effects (SRA) by pointing out the differences with full approach (FRA) whenever they are relevant.

\subsubsection{Augmented waves as a first type of basis functions}\label{aug_fn}

Basis functions of the augmented type are in their own turn subdivided into Linearized Augmented Plane Waves (LAPW) and Augmented Plane Waves (APW). Both LAPW and APW are characterized by relativistic plane waves
(Loucks \cite{pr_139_A1333}) in the interstitial region:
\begin{equation} \label{rpw}
\Pi^{\mathbf{k}}_{\mathbf{G},s}(\mathbf{r})|_{\Omega_{i}}=\frac{N_{\mathbf{k+G}}}{\sqrt{\Omega_{0}}}
\left( \begin{array}{c} u_{s}
\\
\frac{c \mathbf{\sigma} \cdot \left( \mathbf{k+G} \right)}{c^2+
\epsilon_{\mathbf{k+G}}^{+}} u_{s}
\end{array} \right)
\exp[\text{i} (\mathbf{k+G}) \cdot \mathbf{r}],
\end{equation}
where $\mathbf{G}$ are vectors of reciprocal lattice, $\epsilon_{\mathbf{k+G}}^{+}$ is positive relativistic energy associated with wave vector $\mathbf{k+G}$, i.e., $2\epsilon_{\mathbf{k+G}}^{+}=-c^{2}+c
\sqrt{c^{2}+4(\mathbf{k+G})^{2}}$, $N_{\mathbf{k+G}}$ stands for the normalization factor, and $u_{s}$ is spinor function with spin index $s=\pm \frac{1}{2}$. In simplified approach we just neglect by small components in the interstitial region:

\begin{equation} \label{rpw_s}
\Pi^{\mathbf{k}}_{\mathbf{G},s}(\mathbf{r})|_{\Omega_{i}}=\frac{1}{\sqrt{\Omega_{0}}}
\left( \begin{array}{c} u_{s}
\\
0
\end{array} \right)
\exp[\text{i} (\mathbf{k+G}) \cdot \mathbf{r}].
\end{equation}

Inside the MT spheres, the construction is different for LAPW and APW. Let's discuss first the LAPW type. In this case, basis functions inside MT are constructed from solutions of the Dirac equation with spherical-only component of the scalar effective potential $V_{0}^{t}(\mathbf{r})$ and with $\mathbf{B}=0$
\begin{equation} \label{rad1}
[h^{t}(r)-\epsilon_{il}^{t}] \left(
\begin{array}{cc}
g_{il}^{t}(r)  &  \Omega_{i;l;\mu}(\theta, \varphi)  \\
\frac{\text{i}}{c}f_{il}^{t}(r)   &   \Omega_{-i;l+2i;\mu}(\theta,
\varphi)
\end{array}  \right)
=0,
\end{equation}
and from their derivatives with respect to energy, which one finds from the equation
\begin{eqnarray} \label{rad2}
[h^{t}(r)-\epsilon_{il}^{t}] &\left(
\begin{array}{cc}
\dot{g}_{il}^{t}(r)  &  \Omega_{i;l;\mu}(\theta, \varphi)  \\
\frac{\text{i}}{c} \dot{f}_{il}^{t}(r)   &
\Omega_{-i;l+2i;\mu}(\theta, \varphi)
\end{array}  \right)\nonumber\\&
=\left(
\begin{array}{cc}
g_{il}^{t}(r)  &  \Omega_{i;l;\mu}(\theta, \varphi)  \\
\frac{\text{i}}{c}f_{il}^{t}(r)   &   \Omega_{-i;l+2i;\mu}(\theta,
\varphi)
\end{array}  \right).
\end{eqnarray}

In these equations, $h^{t}(r)=\hat{T}+V_{0}^{t}(r)$, and $\epsilon_{il}^{t}$ are energy parameters which one finds self-consistently.
$\Omega_{i;l;\mu}(\theta, \varphi)$ are spin-angular functions, $l$ is orbital momentum, $i$ defines total momentum $j$ via $j=l+i$ ($i=\pm \frac{1}{2}$, not to mess with the complex number $\text{i}$). $\mu$ is $z$-projection of the total momentum. Relativistic quantum number $\kappa$ is related with $l$ and $i$ by the following definitions
\begin{equation} \label{kappa}
\begin{array}{ll}
\kappa=l, & i=-\frac{1}{2} \nonumber \\
\kappa=-l-1, & i=\frac{1}{2}.
\end{array}
\end{equation}

Spin-angular functions are build according to the general scheme of angular momenta coupling:
\begin{equation}\label{spinors}
\Omega_{i;l;\mu}(\theta, \varphi)=\sum_{s=\pm \frac{1}{2}}
C_{is}^{l\mu} Y_{l;\mu-s}(\theta, \varphi)u_{s},
\end{equation}
where $Y_{l;\mu-s}(\theta, \varphi)$ are complex spherical harmonics,
\cite{a69}. $C_{is}^{l;\mu}$ are Clebsch-Gordan coefficients conveniently defined with the use of a parameter $u_{l\mu}=\mu/(l+1/2)$:
\begin{eqnarray} \label{CLGor}
C^{l;\mu}&= \left(
\begin{array}{cc}
C_{-\frac{1}{2} -\frac{1}{2}}^{l;\mu}   & C_{-\frac{1}{2}
\frac{1}{2}}^{l;\mu}  \\
C_{\frac{1}{2} -\frac{1}{2}}^{l;\mu}   & C_{\frac{1}{2}
\frac{1}{2}}^{l;\mu}
\end{array} \right) \nonumber\\&= \frac{1}{\sqrt{2}} \left(
\begin{array}{cc}
\sqrt{1+u_{l\mu}}    & -\sqrt{1-u_{l\mu}}   \\
\sqrt{1-u_{l\mu}}    & \sqrt{1+u_{l\mu}}
\end{array} \right).
\end{eqnarray}

Spin-angular functions are ortho-normalized:
\begin{equation}\label{ortho}
\int \Omega_{i;l;\mu}^{\dag}(\theta, \varphi)
\Omega_{i';l';\mu'}(\theta, \varphi) \sin \theta \text{d} \theta
\text{d} \varphi= \delta_{ii'} \delta_{ll'} \delta_{\mu \mu'}.
\end{equation}
Radial functions $g_{il}^{t}(r)$ and $f_{il}^{t}(r)$ are the solutions to the following system of differential equations:
\begin{eqnarray} \label{difeq}
\frac{\text{d}(rg_{il}^{t})}{\text{d}
r}=-\frac{\kappa}{r}(rg_{il}^{t}) + \frac{c^2 +
\epsilon_{il}^{t}-V_{0}^{t}}{c^2} (rf_{il}^{t})   \\
\frac{\text{d}(rf_{il}^{t})}{\text{d}
r}=\frac{\kappa}{r}(rf_{il}^{t}) - (
\epsilon_{il}^{t}-V_{0}^{t})(rg_{il}^{t}).
\end{eqnarray}
For the energy derivatives the following system is solved
\begin{eqnarray} \label{difeq1}
\frac{\text{d}(r\dot{g_{il}^{t}})}{\text{d}
r}=-\frac{\kappa}{r}(r\dot{g_{il}^{t}}) + \frac{c^2 +
\epsilon_{il}^{t}-V_{0}^{t}}{c^2} (r\dot{f_{il}^{t}}) + \frac{1}{c^2}(rf_{il}^{t})   \\
\frac{\text{d}(r\dot{f_{il}^{t}})}{\text{d}
r}=\frac{\kappa}{r}(r\dot{f_{il}^{t}}) - (
\epsilon_{il}^{t}-V_{0}^{t})(r\dot{g_{il}^{t}})-(rg_{il}^{t}).
\end{eqnarray}
It is better to use normalized radial functions:

\begin{equation} \label{norma}
\langle g^{2}\rangle+\langle f^{2}\rangle/c^{2}=1,
\end{equation}
which means the orthogonality to their energy derivatives:
\begin{equation} \label{orthog}
\langle g|\dot{g}\rangle+\langle f|\dot{f}\rangle/c^{2}=0.
\end{equation}

It is convenient to build two linear combinations $R^{(1)t}_{il \mu}(\mathbf{r})$ and $R^{(2)t}_{il
\mu}(\mathbf{r})$ from the solutions of (\ref{rad1}) and (\ref{rad2}):
\begin{eqnarray} \label{lincomb1}
R^{(1)t}_{il \mu}(\mathbf{r})&= a^{(1)t}_{il} \left(
\begin{array}{cc}
g_{il}^{t}(r)  &  \Omega_{i;l;\mu}(\theta, \varphi)  \\
\frac{\text{i}}{c}f_{il}^{t}(r)   &   \Omega_{-i;l+2i;\mu}(\theta,
\varphi)
\end{array}  \right) \nonumber\\&+ b^{(1)t}_{il}
\left(
\begin{array}{cc}
\dot{g}_{il}^{t}(r)  &  \Omega_{i;l;\mu}(\theta, \varphi)  \\
\frac{\text{i}}{c} \dot{f}_{il}^{t}(r)   &
\Omega_{-i;l+2i;\mu}(\theta, \varphi)
\end{array}  \right),
\end{eqnarray}
and
\begin{eqnarray} \label{lincomb2}
R^{(2)t}_{il \mu}(\mathbf{r})&= a^{(2)t}_{il} \left(
\begin{array}{cc}
g_{il}^{t}(r)  &  \Omega_{i;l;\mu}(\theta, \varphi)  \\
\frac{\text{i}}{c}f_{il}^{t}(r)   &   \Omega_{-i;l+2i;\mu}(\theta,
\varphi)
\end{array}  \right) \nonumber\\&+ b^{(2)t}_{il}
\left(
\begin{array}{cc}
\dot{g}_{il}^{t}(r)  &  \Omega_{i;l;\mu}(\theta, \varphi)  \\
\frac{\text{i}}{c} \dot{f}_{il}^{t}(r)   &
\Omega_{-i;l+2i;\mu}(\theta, \varphi)
\end{array}  \right),
\end{eqnarray}
in a special way which is slightly different for FRA and SRA. In the case of FRA, the following two conditions are met at the boundary of the MT sphere:
\begin{equation} \label{prop1}
R^{(1)t}_{il \mu}(\mathbf{S_{t}})= \left(
\begin{array}{cc}
1  &  \Omega_{i;l;\mu}(\theta, \varphi)  \\
0  &   \Omega_{-i;l+2i;\mu}(\theta, \varphi)
\end{array}  \right),
\end{equation}
and
\begin{equation} \label{prop2}
R^{(2)t}_{il \mu}(\mathbf{S_{t}})= \left(
\begin{array}{cc}
0  &  \Omega_{i;l;\mu}(\theta, \varphi)  \\
\frac{\text{i}}{c} \cdot 1  &   \Omega_{-i;l+2i;\mu}(\theta,
\varphi)
\end{array}  \right).
\end{equation}

In the case of SRA, we impose the conditions on the big component and its radial derivative (instead of small component):

\begin{equation} \label{prop1s}
\begin{aligned}
R^{(1)t}_{il \mu}(\mathbf{S_{t}})= \left(
\begin{array}{cc}
1  &  \Omega_{i;l;\mu}(\theta, \varphi)  \\
\frac{\text{i}}{c} \cdot f^{(1)}  &   \Omega_{-i;l+2i;\mu}(\theta, \varphi)
\end{array}  \right),\\
R'^{(1)t}_{il \mu}(\mathbf{S_{t}})= \left(
\begin{array}{cc}
0  &  \Omega_{i;l;\mu}(\theta, \varphi)  \\
\frac{\text{i}}{c} \cdot f'^{(1)}  &   \Omega_{-i;l+2i;\mu}(\theta, \varphi)
\end{array}  \right),
\end{aligned}
\end{equation}
and
\begin{equation} \label{prop2s}
\begin{aligned}
R^{(2)t}_{il \mu}(\mathbf{S_{t}})= \left(
\begin{array}{cc}
0  &  \Omega_{i;l;\mu}(\theta, \varphi)  \\
\frac{\text{i}}{c} \cdot f^{(2)}  &   \Omega_{-i;l+2i;\mu}(\theta,
\varphi)
\end{array}  \right),\\
R'^{(2)t}_{il \mu}(\mathbf{S_{t}})= \left(
\begin{array}{cc}
1  &  \Omega_{i;l;\mu}(\theta, \varphi)  \\
\frac{\text{i}}{c} \cdot f'^{(2)}  &   \Omega_{-i;l+2i;\mu}(\theta,
\varphi)
\end{array}  \right).
\end{aligned}
\end{equation}

As one can see, in SRA case we do not impose any special conditions on the small components at the sphere boundaries.

Now we are in a position to write down our basis functions of the first type inside the MT spheres as linear combinations of $R^{(1)t}_{il \mu}(\mathbf{r})$ and
$R^{(2)t}_{i l \mu}(\mathbf{r})$
\begin{equation} \label{basmt1}
\Pi^{\mathbf{k}}_{\mathbf{G}s}(\mathbf{r})|_{\Omega_{t}}= e^{i\mathbf{k}\mathbf{t}}\sum_{i l \mu}\sum_{w=1}^{2}
y^{(w)\mathbf{k}}_{til \mu;\mathbf{G}s}R^{(w)t}_{il \mu}(\mathbf{r}).
\end{equation}

Coefficients $y^{(w)\mathbf{k}}_{til \mu;\mathbf{G}s}$ are also different for FRA and SRA. For the FRA case, the following expansion of the relativistic plane wave in spherical spinors is useful in obtaining them:

\begin{widetext}
\begin{align} \label{rpw_exp}
\frac{N_{\mathbf{k+G}}}{\sqrt{\Omega_{0}}}&
\left( \begin{array}{c} u_{s}
\\
\frac{c \mathbf{\sigma} \cdot \left( \mathbf{k+G} \right)}{c^2+
\epsilon_{\mathbf{k+G}}^{+}} u_{s}
\end{array} \right)
\exp[\text{i} (\mathbf{k+G}) \cdot \mathbf{r}]\nonumber\\=&\frac{N_{\mathbf{k+G}}}{\sqrt{\Omega_{0}}}
\sum_{i l \mu}4 \pi \text{i}^{l} 
C_{is}^{l \mu} Y_{l \mu-s}^{*}(\widehat{\mathbf{k+G}})\left(
\begin{array}{cc}
j_{l}(|\mathbf{k+G}|r)  &  \Omega_{i;l;\mu}(\theta, \varphi)  \\
\frac{\text{i}}{c} \nu_{i}\frac{c^{2} \mathbf{\sigma} \cdot \left( \mathbf{k+G} \right)}{c^2+
\epsilon_{\mathbf{k+G}}^{+}}j_{l+2i}(|\mathbf{k+G}|r)   &   \Omega_{-i;l+2i;\mu}(\theta,
\varphi)
\end{array}  \right),
\end{align}
\end{widetext}
with $\nu_{i}=1$ for $i=-1/2$ and $\nu_{i}=-1$ for $i=1/2$, which one can combine in $\nu_{i}=-2i$.

The coefficients, correspondingly, are obtained from (\ref{rpw_exp}), the conditions (\ref{prop1}, \ref{prop2}), and the continuity requirement for the big and small components of functions, defined in (\ref{rpw}) and (\ref{basmt1}):

\begin{align} \label{defy}
y^{(1)\mathbf{k}}_{til \mu;\mathbf{G}s}&=\frac{4 \pi}{\sqrt{\Omega}}
\text{i}^{l} \text{e}^{\text{i}\mathbf{G}\mathbf{t}} N_{\mathbf{k+G}}
C_{is}^{l \mu}\nonumber\\&\times j_{l}(|\mathbf{k+G}|S_{t}) Y_{l \mu-s}^{*}(\widehat{\mathbf{k+G}})  \nonumber \\
y^{(2)\mathbf{k}}_{til \mu;\mathbf{G}s}&=\frac{4 \pi}{\sqrt{\Omega}}
\text{i}^{l-2i} \text{e}^{\text{i}\mathbf{G}\mathbf{t}}
N_{\mathbf{k+G}} C_{is}^{l \mu} \nonumber\\&\times\frac{c^{2}|\mathbf{k+G}|}{c^2+\epsilon_{\mathbf{k+G}}^{+}} j_{l+2i}(|\mathbf{k+G}|S_{t}) Y_{l\mu-s}^{*}(\widehat{\mathbf{k+G}}).
\end{align}

In SRA case, we require the continuity of the big component and its derivative at the MT boundary. So, the simplified expansion of the relativistic plane wave is used:
\begin{align} \label{rpw_exps}
\frac{1}{\sqrt{\Omega_{0}}}&
\left( \begin{array}{c} u_{s}
\\
0
\end{array} \right)
\exp[\text{i} (\mathbf{k+G}) \cdot \mathbf{r}]\nonumber\\=&\frac{1}{\sqrt{\Omega_{0}}}
\sum_{i l \mu}4 \pi \text{i}^{l} 
C_{is}^{l \mu} Y_{l \mu-s}^{*}(\widehat{\mathbf{k+G}})\nonumber\\&\times\left(
\begin{array}{cc}
j_{l}(|\mathbf{k+G}|r)  &  \Omega_{i;l;\mu}(\theta, \varphi)  \\
0   &   
\end{array}  \right).
\end{align}

The coefficients are obtained from (\ref{rpw_exps}), the conditions (\ref{prop1s}, \ref{prop2s}), and the continuity requirement for the big component and its radial derivative:

\begin{align} \label{defys}
y^{(1)\mathbf{k}}_{til \mu;\mathbf{G}s}=\frac{4 \pi}{\sqrt{\Omega}}
\text{i}^{l} \text{e}^{\text{i}\mathbf{G}\mathbf{t}} 
C_{is}^{l \mu} j_{l}(|\mathbf{k+G}|S_{t}) Y_{l \mu-s}^{*}(\widehat{\mathbf{k+G}})  \nonumber \\
y^{(2)\mathbf{k}}_{til \mu;\mathbf{G}s}=\frac{4 \pi}{\sqrt{\Omega}}
\text{i}^{l} \text{e}^{\text{i}\mathbf{G}\mathbf{t}}
C_{is}^{l \mu} j'_{l}(|\mathbf{k+G}|S_{t}) Y_{l
\mu-s}^{*}(\widehat{\mathbf{k+G}}).
\end{align}

To simplify the notations, it is convenient to combine the definitions of the augmentation coefficients for FRA and SRA. For this purpose, we introduce functions $\overline{j}^{(1)}_{il}$ and $\overline{j}^{(2)}_{il}$ according to the following definitions:

\begin{align} \label{defys1}
\begin{aligned}
\overline{j}^{(1)}_{il}(|\mathbf{k+G}|S_{t})=N_{\mathbf{k+G}} j_{l}(|\mathbf{k+G}|S_{t})\\
\overline{j}^{(2)}_{il}(|\mathbf{k+G}|S_{t})=N_{\mathbf{k+G}} \nu_{i} \frac{c^{2} \mathbf{k+G}}
{c^2+\epsilon_{\mathbf{k+G}}^{+}} j_{l+2i}(|\mathbf{k+G}|S_{t}),
\end{aligned}
\end{align}
for FRA case, and
\begin{align} \label{defys2}
\begin{aligned}
\overline{j}^{(1)}_{il}(|\mathbf{k+G}|S_{t})=j_{l}(|\mathbf{k+G}|S_{t})\\
\overline{j}^{(2)}_{il}(|\mathbf{k+G}|S_{t})=j'_{l}(|\mathbf{k+G}|S_{t})
\end{aligned}
\end{align}
for SRA case.

With these definitions, the equations (\ref{defy}) and (\ref{defys}) can be combined in one:

\begin{align} \label{defyc}
y^{(w)\mathbf{k}}_{til \mu;\mathbf{G}s}=\frac{4 \pi}{\sqrt{\Omega}} \text{i}^{l} 
\text{e}^{\text{i}\mathbf{G}\mathbf{t}} 
C_{is}^{l \mu} \overline{j}^{(w)}_{il}(|\mathbf{k+G}|S_{t}) Y_{l \mu-s}^{*}(\widehat{\mathbf{k+G}}).
\end{align}

In case of APW, the only augmentation requirement is the continuity of the big component at the MT boundary. For this, we need only the solutions to the Eq.(\ref{rad1}). Thus, for the APW type of augmentation the equation (\ref{basmt1}) has only $w=1$ term in the sum. 

At this point it is a good time to point out that LAPW and APW augmentations can be combined flexibly. Namely, in the sum over $l$ in eq.(\ref{basmt1}), one can use APW augmentation for some $l$'s and LAPW augmentation for the rest. Equation (\ref{basmt1}) can, correspondingly, be generalized by introducing the upper limit $N^{t}_{l}=1$ (for those $tl$ where APW augmentation is used) and $N^{t}_{l}=2$ (for those $tl$ where LAPW augmentation is used):

\begin{equation} \label{basmt_gen}
\Pi^{\mathbf{k}}_{\mathbf{G}s}(\mathbf{r})|_{\Omega_{t}}= e^{i\mathbf{k}\mathbf{t}}\sum_{i l \mu}\sum_{w=1}^{N^{t}_{l}}
y^{(w)\mathbf{k}}_{il \mu t;\mathbf{G}s}R^{(w)t}_{il \mu}(\mathbf{r}).
\end{equation}

The function $R^{(w=1)t}_{il \mu}(\mathbf{r})$ for the APW augmentation is found from the corresponding augmentation constraint:
\begin{equation} \label{lincomb_apw}
R^{(1)t}_{il \mu}(\mathbf{r})= \frac{1}{g_{il}^{t}(S_{t})} \left(
\begin{array}{cc}
g_{il}^{t}(r)  &  \Omega_{i;l;\mu}(\theta, \varphi)  \\
\frac{\text{i}}{c}f_{il}^{t}(r)   &   \Omega_{-i;l+2i;\mu}(\theta,
\varphi)
\end{array}  \right).
\end{equation}

The APW augmentation is less restrictive as compared to the LAPW augmentation and, as a result, plane wave expansion in the interstitial region converges faster if the APW augmentation is used \cite{prb_64_195134}. This enhanced efficiency in the interstitial region has, however, a price. The fact, that in APW case only the solution of radial equations is used but not its energy derivative, makes APW basis set less flexible inside the MT spheres as compared to the LAPW case. Madsen et al. \cite{prb_64_195134} proposed to supplement APW with a special combination of solutions and their energy derivatives (we will use common term for them - lo) which is not zero only inside the MT spheres and, as it was shown in Ref. \cite{prb_64_195134}, is very efficient in making this extended basis set (APW+lo) even more flexible inside the MT than the LAPW. The definition of these local orbitals in FRA/SRA case, together with other types of local orbitals, is described in the next section.

\subsubsection{Local functions as a second type of basis functions}\label{loc_fn}

Basis functions of the second type are local functions. They are defined only inside the MT spheres and can be used to enhance the variational freedom for the valence (conduction) states and to describe the semicore states on the same footing as the valence states. Let us begin with local orbitals (lo) which were briefly introduced in the previous section. In this case, the combination of the solutions of radial equations and their energy derivatives is constructed:

\begin{eqnarray} \label{comb_lo}
R^{(lo)t}_{il \mu}(\mathbf{r})&= a^{(lo)t}_{il} \left(
\begin{array}{cc}
g_{il}^{t}(r)  &  \Omega_{i;l;\mu}(\theta, \varphi)  \\
\frac{\text{i}}{c}f_{il}^{t}(r)   &   \Omega_{-i;l+2i;\mu}(\theta,
\varphi)
\end{array}  \right) \nonumber\\&+ b^{(lo)t}_{il}
\left(
\begin{array}{cc}
\dot{g}_{il}^{t}(r)  &  \Omega_{i;l;\mu}(\theta, \varphi)  \\
\frac{\text{i}}{c} \dot{f}_{il}^{t}(r)   &
\Omega_{-i;l+2i;\mu}(\theta, \varphi)
\end{array}  \right),
\end{eqnarray}
where the coefficients $a^{(lo)t}_{il}$ and $b^{(lo)t}_{il}$ are defined from a condition that the combination has zero big component at the MT boundary and normalized. The solutions and their energy derivatives are found from the same equations (\ref{rad1})-(\ref{rad2}) and with the same energy parameters $\epsilon_{il}^{t}$ as augmentation functions in the LAPW/APW case. It is important to mention, that the above construction of lo's leaves their small components (FRA) or their radial derivatives (SRA) arbitrary. This fact brings in additional discontinuities in the basis set which will be discussed later in the section devoted to the evaluation of matrix elements.

Next type of local functions is High Derivative Local Orbitals (HDLO). They are defined as the following linear combination
\begin{eqnarray} \label{lincomb3}
R^{(HDLO)t}_{il\mu}(\mathbf{r})&= a^{(HDLO)t}_{il\mu} \left(
\begin{array}{cc}
g_{il}^{t}(r)  \Omega_{i;l;\mu}(\theta, \varphi)  \\
\frac{\text{i}}{c}f_{il}^{t}(r)   \Omega_{-i;l+2i;\mu}(\theta,
\varphi)
\end{array}  \right) \nonumber\\&+ b^{(HDLO)t}_{il\mu}
\left(
\begin{array}{cc}
\dot{g}_{il}^{t}(r)    \Omega_{i;l;\mu}(\theta, \varphi)  \\
\frac{\text{i}}{c} \dot{f}_{il}^{t}(r)   
\Omega_{-i;l+2i;\mu}(\theta, \varphi)
\end{array}  \right) \nonumber \\&+c^{(HDLO)t}_{il\mu}
\left(
\begin{array}{cc}
\ddot{g}_{il}^{t}(r)    \Omega_{i;l;\mu}(\theta, \varphi)  \\
\frac{\text{i}}{c} \ddot{f}_{il}^{t}(r)   
\Omega_{-i;l+2i;\mu}(\theta, \varphi)
\end{array}  \right),
\end{eqnarray}
with the requirement that both big and small components (big component and its derivative in SRA) of this combination are zero at the MT boundary and the combination is normalized. One can show that in FRA case the coefficient $b^{(HDLO)t}_{il\mu}$ is identically zero. In SRA case, however, it is small but not zero. So we keep it for generality. Second derivatives are obtained as solutions to the following system:
\begin{eqnarray} \label{rad3}
[h^{t}(r)-\epsilon_{il}^{t}]& \left(
\begin{array}{cc}
\dot{g}_{il}^{t}(r)    \Omega_{i;l;\mu}(\theta, \varphi)  \\
\frac{\text{i}}{c} \dot{f}_{il}^{t}(r)   
\Omega_{-i;l+2i;\mu}(\theta, \varphi)
\end{array}  \right)\nonumber\\&
=2\left(
\begin{array}{cc}
g_{il}^{t}(r)    \Omega_{i;l;\mu}(\theta, \varphi)  \\
\frac{\text{i}}{c}f_{il}^{t}(r)      \Omega_{-i;l+2i;\mu}(\theta,
\varphi)
\end{array}  \right).
\end{eqnarray}
with the radial components being the solutions to the following equations:
\begin{eqnarray} \label{difeq2}
\begin{aligned}
\frac{\text{d}(r\dot{g_{il}^{t}})}{\text{d}
r}=-\frac{\kappa}{r}(r\dot{g_{il}^{t}}) + \frac{c^2 +
\epsilon_{il}^{t}-V_{0}^{t}}{c^2} (r\dot{f_{il}^{t}}) + \frac{1}{c^2}(rf_{il}^{t})   \\
\frac{\text{d}(r\dot{f_{il}^{t}})}{\text{d}
r}=\frac{\kappa}{r}(r\dot{f_{il}^{t}}) - (
\epsilon_{il}^{t}-V_{0}^{t})(r\dot{g_{il}^{t}})-2(rg_{il}^{t}).
\end{aligned}
\end{eqnarray}

High Energy Local Orbitals (HELO) are defined exactly as HDLO with the only difference that the solutions of radial equations, together with their first and second derivatives entering the linear combination, are all found with energy parameters $\epsilon_{nil}^{t}$ corresponding to high energy states or to the semicore energy.
In order to distinguish HELO from HDLO, we use additional index $n$ for HELO's which specifies principal quantum number. One has to mention that the definition of HELO's in this work is slightly different from the construction in Refs. \cite{cpc_184_2670,cpc_220_230}. Namely, the HELO's commonly are constructed as linear combinations of the solution and its first derivative found at the same $\epsilon_{il}^{t}$ as the augmentation functions, and the solution found at the high energy $\epsilon_{nil}^{t}$. Both variants have been implemented in the FlapwMBPT code. A few tests have shown very little difference in performance. Thus, it was decided to use in this work the implementation introduced above (with all functions found with the same high energy) and to use abbreviation HELO for it.

For any type of local orbitals (LOC) we use formal Bloch's sums of the functions $R^{(LOC)t}_{nil\mu}$
\begin{equation} \label{basmt2}
\Lambda^{\mathbf{k}}_{tnil\mu}(\mathbf{r})=\sum_{\mathbf{R}}
\text{e}^{\text{i}\mathbf{k(t+R)}} R^{(LOC)t}_{nil\mu}(\mathbf{r}_{t}),
\end{equation}
as the basis functions of the second type. Here $\mathbf{R}$ are lattice translation vectors and 
$\mathbf{r}_{t}$ stands for the vector measured from the center of atom $t$: $\mathbf{r}_{t}=\mathbf{r}-\mathbf{t}$.

Now one can write down the solution of the Dirac-Kohn-Sham equations (\ref{ksh}) as linear combination of the functions of the first and the second type:
\begin{equation} \label{wf}
\Psi^{\mathbf{k}}_{\lambda}(\mathbf{r})=\sum_{\mathbf{G} s} A^{\mathbf{k}}_{\mathbf{G}s;\lambda}
\Pi^{\mathbf{k}}_{\mathbf{G}s}(\mathbf{r}) +\sum_{t nil\mu} B^{\mathbf{k}}_{tnil\mu;\lambda}
\Lambda^{\mathbf{k}}_{tnil\mu}(\mathbf{r}).
\end{equation}
Coefficients $A^{\mathbf{k}}_{\mathbf{G} s;\lambda}$ и $B^{\mathbf{k}}_{tnil\mu;\lambda}$ are to be found variationally.

\subsection{Matrix elements of hamiltonian and overlap matrices}

Solution of the Dirac-Kohn-Sham equations (\ref{ksh}) for solid with the use of representation (\ref{wf}) leads to the generalized eigen value problem:
\begin{equation} \label{geigen}
\left[
\begin{array}{cc}
H_{AA}   &    H_{BA}^{\dag}  \\
H_{BA}   &    H_{BB}
\end{array}  \right]
\left|
\begin{array}{c}
A  \\
B
\end{array} \right| = E
\left[
\begin{array}{cc}
O_{AA}   &    O_{BA}^{\dag}  \\
O_{BA}   &    O_{BB}
\end{array}  \right]
\left|
\begin{array}{c}
A  \\
B
\end{array} \right|,
\end{equation}
where the indexes were discarded for brevity. Matrix elements of the matrices $O$ and $H$
(overlap and hamiltonian correspondingly) can be divided into interstitial and MT components. Furthermore, the MT component can be divided into spherical part (matrix elements of kinetic energy and $l=0$ part of the scalar effective potential) and non-spherical part which includes full magnetic field contribution.

\subsubsection{Interstitial part of the matrix elements of H and O}
\label{ME_int}

Interstitial part contributes only to the matrix elements between functions of the first type.
Its evaluation in FRA case is based on the definition of basis functions in the interstitial region (\ref{rpw}) and properties of the Pauli matrices. Final result for the overlap integral:

\begin{widetext}
\begin{align}\label{oint}
O_{\mathbf{G}s;\mathbf{G'}s'}^{\mathbf{k}}|_{\Omega_{I}} =
\frac{N_{\mathbf{k+G}}N_{\mathbf{k+G'}}}{\Omega_{0}} \int_{\Omega_{I}}
\text{e}^{\text{i}(\mathbf{G'-G})\mathbf{r}} \text{d} \mathbf{r} \left\{
\delta_{ss'} +
\frac{c^2}{(c^2+\epsilon_{\mathbf{k+G}}^{+})(c^2+\epsilon_{\mathbf{k+G'}}^{+})}\right.
\nonumber  \\
\left. \times \left[ (\mathbf{k+G}) \cdot
(\mathbf{k+G'})\delta_{ss'}+\text{i}[(\mathbf{k+G}) \times (\mathbf{k+G'})]
\cdot <u_{s}|\boldsymbol{\sigma}|u_{s'}> \right] \right\}.
\end{align}
\end{widetext}

In the evaluation of matrix elements of the kinetic energy operator, the average should be taken (as it follow from Eq.(\ref{sm_rl1})).

\begin{equation}\label{kinint}
K_{\mathbf{G}s;\mathbf{G'}s'}^{\mathbf{k}}|_{\Omega_{I}} =
\frac{1}{2}\left(\epsilon_{\mathbf{k+G}}^{+}+\epsilon_{\mathbf{k+G'}}^{+}\right)
O_{\mathbf{G}s;\mathbf{G'}s'}^{\mathbf{k}}|_{\Omega_{I}}.
\end{equation}

Scalar potential and magnetic field matrix elements:

\begin{widetext}
\begin{eqnarray}\label{vint}
V_{\mathbf{G}s;\mathbf{G'}s'}^{\mathbf{k}}|_{\Omega_{I}} =
\frac{N_{\mathbf{k+G}}N_{\mathbf{k+G'}}}{\Omega_{0}} \int_{\Omega_{I}}
V_{eff}(\mathbf{r}) \text{e}^{\text{i}(\mathbf{G'-G})\mathbf{r}} \text{d} \mathbf{r}
\{ \delta_{ss'} +
\frac{c^2}{(c^2+\epsilon_{\mathbf{k+G}}^{+})(c^2+\epsilon_{\mathbf{k+G'}}^{+})}
\nonumber  \\
\times \left[ (\mathbf{k+G}) \cdot
(\mathbf{k+G'})\delta_{ss'}+\text{i}[(\mathbf{k+G}) \times (\mathbf{k+G'})]
\cdot <u_{s}|\boldsymbol{\sigma}|u_{s'}> \right] \}.
\end{eqnarray}

\begin{eqnarray}\label{bint}
B_{\mathbf{G}s;\mathbf{G'}s'}^{\mathbf{k}}|_{\Omega_{I}} =
\frac{N_{\mathbf{k+G}}N_{\mathbf{k+G'}}}{\Omega_{0}} \int_{\Omega_{I}}
B_{eff}(\mathbf{r}) \text{e}^{\text{i}(\mathbf{G'-G})\mathbf{r}} \text{d} \mathbf{r}
\{ \mathbf{n}_{B} \cdot
<u_{s}|\boldsymbol{\sigma}|u_{s'}>  \nonumber \\
-
\frac{c^2}{(c^2+\epsilon_{\mathbf{k+G}}^{+})(c^2+\epsilon_{\mathbf{k+G'}}^{+})}
[ (\mathbf{n}_{B} \cdot (\mathbf{k+G})) ((\mathbf{k+G'})
\cdot <u_{s}|\boldsymbol{\sigma}|u_{s'}>) \nonumber \\
+ (\mathbf{n}_{B} \cdot (\mathbf{k+G'}))
 ((\mathbf{k+G}) \cdot
<u_{s}|\boldsymbol{\sigma}|u_{s'}>) \nonumber \\
- ((\mathbf{k+G}) \cdot (\mathbf{k+G'})) (\mathbf{n}_{B} \cdot
<u_{s}|\boldsymbol{\sigma}|u_{s'}>) \nonumber \\
- \text{i}\mathbf{n}_{B} \cdot [(\mathbf{k+G}) \times (\mathbf{k+G'})]
\delta_{ss'} ] \},
\end{eqnarray}
\end{widetext}
where vector $\mathbf{n}_{B}$ indicates the direction of the magnetic field.

In the SRA case, the formulae are considerably simpler:

Overlap integral
\begin{align}\label{oint_s}
O_{\mathbf{G}s;\mathbf{G'}s'}^{\mathbf{k}}|_{\Omega_{I}} =
\frac{\delta_{ss'}}{\Omega_{0}} \int_{\Omega_{I}}
\text{e}^{\text{i}(\mathbf{G'-G})\mathbf{r}} \text{d} \mathbf{r}.
\end{align}

Kinetic energy
\begin{equation}\label{kinint_s}
K_{\mathbf{G}s;\mathbf{G'}s'}^{\mathbf{k}}|_{\Omega_{I}} =
\frac{1}{2}\left(|\mathbf{k+G}|^{2}+|\mathbf{k+G'}|^{2}\right)
O_{\mathbf{G}s;\mathbf{G'}s'}^{\mathbf{k}}|_{\Omega_{I}}.
\end{equation}

Scalar potential matrix elements

\begin{eqnarray}\label{vint_s}
V_{\mathbf{G}s;\mathbf{G'}s'}^{\mathbf{k}}|_{\Omega_{I}} =
\frac{\delta_{ss'}}{\Omega_{0}} \int_{\Omega_{I}}
V_{eff}(\mathbf{r}) \text{e}^{\text{i}(\mathbf{G'-G})\mathbf{r}} \text{d} \mathbf{r}.
\end{eqnarray}

Magnetic field matrix elements

\begin{eqnarray}\label{bint_s}
B_{\mathbf{G}s;\mathbf{G'}s'}^{\mathbf{k}}|_{\Omega_{I}} =
\frac{1}{\Omega_{0}} \int_{\Omega_{I}}
B_{eff}(\mathbf{r}) \text{e}^{\text{i}(\mathbf{G'-G})\mathbf{r}} \text{d} \mathbf{r}
\mathbf{n}_{B} \cdot
<u_{s}|\boldsymbol{\sigma}|u_{s'}>.
\end{eqnarray}

\subsubsection{MT part of the matrix elements of overlap and spherical part of hamiltonian} \label{MT_sph}

Overlap matrix and the matrix of the spherical part of the hamiltonian in MT spheres are diagonal in indexes ($il\mu$). Let us consider first the AA block of the matrices which is formed by two augmentation functions. Using representation (\ref{basmt_gen}) one obtains the volume integral contribution to the hamiltonian matrix:

\begin{align}\label{hmtaa}
H_{\mathbf{G}s;\mathbf{G'}s'}^{\mathbf{k}} = \sum_{t}
\sum_{il\mu} \sum_{(ww')=1}^{N^{t}_{l}}y^{^{*}(w)\mathbf{k}}_{til\mu;\mathbf{G}s}
y^{(w')\mathbf{k}}_{til\mu;\mathbf{G}'s'}h^{til}_{ww'},
\end{align}
with $h^{til}_{ww'}=\frac{1}{2}\left(\langle R^{(w)t}_{il}|h^{t}|R^{(w')t}_{il}\rangle_{t}+\langle R^{(w')t}_{il}|h^{t}|R^{(w)t}_{il}\rangle_{t}\right)$, and the contribution to the overlap matrix
\begin{align}\label{omtaa}
O_{\mathbf{G}s;\mathbf{G'}s'}^{\mathbf{k}} = \sum_{t}
\sum_{il\mu} \sum_{(ww')=1}^{N^{t}_{l}}y^{^{*}(w)\mathbf{k}}_{til\mu;\mathbf{G}s}
y^{(w')\mathbf{k}}_{til\mu;\mathbf{G}'s'}o^{til}_{ww'},
\end{align}
with $o^{til}_{ww'}=\langle R^{(w)t}_{il}|R^{(w')t}_{il}\rangle_{t}$.

In these formulae, the integrals $h^{til}_{ww'}$ and $o^{til}_{ww'}$ do not depend on the projection of total momentum $\mu$. So, one can sum over the projection and do it analytically with the help of the following identity (can be proved based on the explicit form of Clebsch-Gordan coefficients \eqref{CLGor} and addition theorem for the spherical harmonics):
\begin{widetext}
\begin{align}\label{sum_y}
\sum_{\mu}C^{l\mu}_{is}C^{l\mu}_{is'}Y_{l\mu-s}(\widehat{\mathbf{k+G}})Y^{*}_{l\mu-s'}(\widehat{\mathbf{k+G'}})
=
\frac{|\kappa|}{4\pi}P_{l}(\widehat{\mathbf{k+G;k+G'}})\delta_{ss'}
\nonumber\\
+\frac{iS_{\kappa}}{4\pi}P'_{l}(\widehat{\mathbf{k+G;k+G'}})
\frac{\left[(\mathbf{k+G})\times(\mathbf{k+G'})\right]\cdot\langle
s|\mathbf{\sigma}|s'\rangle}{|\mathbf{k+G}|\cdot|\mathbf{k+G'}|},
\end{align}
\end{widetext}
where the relativistic quantum number $\kappa=l(l+1)-j(j+1)-1/4$ and
$S_{\kappa}$ is the sign of $\kappa$. Denoting this sum as $\frac{1}{4\pi}D^{\mathbf{k}il}_{\mathbf{G}s;\mathbf{G}'s'}$ and using the definition (\ref{defyc}) we can perform the summation over projections of the total momentum in (\ref{hmtaa}) and (\ref{omtaa}):

\begin{align}\label{hmtaa1}
&H_{\mathbf{G}s;\mathbf{G'}s'}^{\mathbf{k}} = \frac{4\pi}{\Omega_{0}}\sum_{t}e^{i(\mathbf{G}'-\mathbf{G})\mathbf{t}}
\sum_{il} D^{\mathbf{k}il}_{\mathbf{G}s;\mathbf{G}'s'}\nonumber\\&\times\sum_{(ww')=1}^{N^{t}_{l}}\overline{j}^{(w)}_{il}(|\mathbf{k+G}|S_{t})
\overline{j}^{(w')}_{il}(|\mathbf{k+G'}|S_{t})h^{til}_{ww'}\nonumber\\&=\sum_{t}F^{t}_{\mathbf{G'-G}}\sum_{il}D^{\mathbf{k}il}_{\mathbf{G}s;\mathbf{G}'s'}\overline{h}^{til}_{\mathbf{GG}'},
\end{align}
with 
\begin{eqnarray}
\begin{aligned}
F^{t}_{\mathbf{G'-G}}=\frac{4\pi}{\Omega_{0}}e^{i(\mathbf{G}'-\mathbf{G})},\\
\overline{h}^{til}_{\mathbf{GG}'}=\sum_{(ww')=1}^{N^{t}_{l}}\overline{j}^{(w)}_{il}(|\mathbf{k+G}|S_{t})
\overline{j}^{(w')}_{il}(|\mathbf{k+G'}|S_{t})h^{til}_{ww'},
\end{aligned}
\end{eqnarray}

and, similarly,

\begin{align}\label{omtaa1}
O_{\mathbf{G}s;\mathbf{G'}s'}^{\mathbf{k}} =\sum_{t}F^{t}_{\mathbf{G'-G}}\sum_{il}D^{\mathbf{k}il}_{\mathbf{G}s;\mathbf{G}'s'}\overline{o}^{til}_{\mathbf{GG}'},
\end{align}
with 
\begin{equation}
\overline{o}^{til}_{\mathbf{GG}'}=\sum_{(ww')=1}^{N^{t}_{l}}\overline{j}^{(w)}_{il}(|\mathbf{k+G}|S_{t})
\overline{j}^{(w')}_{il}(|\mathbf{k+G'}|S_{t})o^{til}_{ww'}.
\end{equation}

Generally, for AA block of the hamiltonian matrix there is also surface contribution as it follows from (\ref{apw_apw_t}). Looking at the structure of the surface terms, one can realize that their contribution can be combined with the volume contribution by modifying the quantities $\overline{h}^{til}_{\mathbf{GG}'}$:

\begin{align}\label{hmtaa2}
&\overline{h}^{til}_{\mathbf{GG}'}\rightarrow \nonumber\\&\overline{h}^{til}_{\mathbf{GG}'}+\frac{S^{2}_{t}}{2}\{2f^{t}_{il}(S_{t})\overline{j}^{(1)}_{il}(|\mathbf{k+G}|S_{t})\overline{j}^{(1)}_{il}(|\mathbf{k+G'}|S_{t})\nonumber\\&-\overline{j}^{(1)}_{il}(|\mathbf{k+G}|S_{t})\overline{j}^{(2)}_{il}(|\mathbf{k+G'}|S_{t})\nonumber\\&-\overline{j}^{(2)}_{il}(|\mathbf{k+G}|S_{t})\overline{j}^{(1)}_{il}(|\mathbf{k+G'}|S_{t})\}.
\end{align}

This modification only exists for those $tl$ for which the augmentation is performed via APW scheme. For SRA case, the small component $f^{t}_{il}(S_{t})$ should be replaced with the radial derivative of the big component.

Let us now consider the contribution to BA block. Using representation (\ref{basmt2}) and representation (\ref{basmt_gen}) one obtains the volume integral contribution to the hamiltonian matrix:

\begin{align}\label{hmtba}
H_{tnil\mu;\mathbf{G'}s'}^{\mathbf{k}} = \sum_{w'=1}^{N^{t}_{l}}
y^{(w')\mathbf{k}}_{til\mu;\mathbf{G}'s'}h^{til}_{nw'},
\end{align}
with 
\begin{equation}
h^{til}_{nw'}=\frac{1}{2}\left(\langle R^{(LOC)t}_{nil}|h^{t}|R^{(w')t}_{il}\rangle_{t}+\langle R^{(w')t}_{il}|h^{t}|R^{(LOC)t}_{nil}\rangle_{t}\right),
\end{equation}
and the contribution to the overlap matrix
\begin{align}\label{omtba}
O_{tnil\mu;\mathbf{G'}s'}^{\mathbf{k}} = \sum_{w'=1}^{N^{t}_{l}}
y^{(w')\mathbf{k}}_{til\mu;\mathbf{G}'s'}o^{til}_{nw'},
\end{align}
with $o^{til}_{nw'}=\langle R^{(LOC)t}_{nil}|R^{(w')t}_{il}\rangle_{t}$.

In the above formulae, LOC can be any of 'lo', 'HDLO', or 'HELO'. Similar to the consideration of the AA block, there is a surface contribution in case LOC='lo'. Surface contribution can be combined with volume contribution by modifying the quantities $h^{til}_{n1}$:

\begin{align}\label{hmtba2}
h^{til}_{n1}\rightarrow h^{til}_{n1}+\frac{S^{2}_{t}}{2}f^{t}_{nil}(S_{t}).
\end{align}

Finally, contribution to the block BB comes only from the volume integrals:

\begin{align}\label{hmtbb}
H_{tnil\mu;tn'il\mu}^{\mathbf{k}} =\frac{1}{2}&
\Big[(\langle R^{(LOC)t}_{nil}|h^{t}|R^{(LOC)t}_{n'il}\rangle_{t}\nonumber\\&
+\langle R^{(LOC)t}_{n'il}|h^{t}|R^{(LOC)t}_{nil}\rangle_{t}\Big],
\end{align}
and
\begin{align}\label{omtbb}
O_{tnil\mu;tn'il\mu}^{\mathbf{k}} =\langle R^{(LOC)t}_{nil}|R^{(LOC)t}_{n'il}\rangle_{t}.
\end{align}

\subsubsection{Matrix elements of the non-spherical part of effective scalar potential and magnetic field in MT spheres} \label{MT_nsph}

The contribution from non-spherical effective potential and from the full magnetic field is evaluated in accordance with the following formulae:
\begin{align}\label{hnmtaa}
&H_{\mathbf{G}s;\mathbf{G'}s'}(\mathbf{k})|_{NMT} = \nonumber\\&\sum_{t} \sum_{il\mu
;i'l'\mu'}\sum_{(ww')=1}^{N^{t}_{l}}y^{^{*}(w)\mathbf{k}}_{til\mu;\mathbf{G}s}
y^{(w')\mathbf{k}}_{ti'l'\mu';\mathbf{G'}s'} \nonumber\\&\times\int_{\Omega_{t}}
R^{^{\dagger}(w)t}_{il\mu}(\mathbf{r}) \hat{H}_{NMT}
R^{(w')t}_{i'l'\mu'}(\mathbf{r}) \text{d} \mathbf{r}.
\end{align}

\begin{eqnarray}\label{hnmtba}
&H_{tnil\mu;\mathbf{G'}s'}(\mathbf{k})|_{NMT} = \sum_{i'l'\mu'}\sum_{w'=1}^{N^{t}_{l}}
y^{(w')\mathbf{k}}_{ti'l'\mu';\mathbf{G'}s'}\nonumber\\&\times \int_{\Omega_{t}}
R^{^{\dag}(LOC)t}_{nil\mu}(\mathbf{r}) \hat{H}_{NMT}
R^{(w')t}_{i'l'\mu'}(\mathbf{r}) \text{d} \mathbf{r}.
\end{eqnarray}

\begin{eqnarray}\label{hnmtbb}
&H_{tnil\mu;t'n'i'l'\mu'}(\mathbf{k})|_{NMT} = \nonumber\\&\delta_{tt'}\int_{\Omega_{t}}
R^{^{\dag}(LOC)t}_{nil\mu}(\mathbf{r}) \hat{H}_{NMT}
R^{(LOC)t}_{n'i'l'\mu'}(\mathbf{r}) \text{d} \mathbf{r}.
\end{eqnarray}

In the above equations the integrals are evaluated with the use of the representation:
\begin{eqnarray}\label{vbharm}
H_{NMT}(\mathbf{r})= \sum_{l\neq0 m} V^{eff}_{lm}(r)\overline{Y}_{lm}(\widehat{\mathbf{r}})
+ \sum_{lm} \mathbf{B}^{eff}_{lm}(r)\overline{Y}_{lm}(\widehat{\mathbf{r}}),
\end{eqnarray}
and definitions (\ref{lincomb1}, \ref{lincomb2}, \ref{lincomb3}) and the identity (\ref{spinors}).

\section{Performance tests}
\label{conv}

\begin{table}[t]
\caption{Structural parameters of the solids considered in this work. Lattice parameters are in Angstroms, MT radii are in atomic units (1 Bohr radius), and the atomic positions are relative to the three primitive translation vectors. If a certain parameter is changing in a specific test, it is detailed in the text.} \label{list_s}
\small
\begin{center}
\begin{tabular}{@{}c c c c c c c} &Space&&&&Atomic&\\
Solid &group&a&b&c&positions&$R_{MT}$\\
\hline\hline
$\alpha$-U&63 &2.854 &5.869 &4.955&0;0.1025;0.25  &2.602333\\
HgSe&216 &6.0854 & &&Hg: 0;0;0&Hg, Se: 2.3113\\
& & & &&Se: 1/4;1/4;1/4  &\\
HgTe&216 &6.4588 & &&Hg: 0;0;0&Hg, Te: 2.4531\\
& & & &&Te: 1/4;1/4;1/4  &\\
FePt&123 &2.7248 & &3.78&Fe: 0;0;0&Fe, Pt: 2.55\\
& & & &&Pt: 1/2;1/2;1/2  &\\
Th&225 &5.0842 & && 0;0;0&3.39685\\
\end{tabular}
\end{center}
\end{table}

\begin{table}[b]
\caption{Principal set up parameters of the studied solids are given. If a certain parameter is changing, it is detailed in the text.} \label{setup_s}
\small
\begin{center}
\begin{tabular}{@{}c c c c c c} &Core&&$L_{max}$&$L_{max}$&\\
Solid &states&Semicore&$\Psi/\rho,V$&APW+lo & $RK_{max}$ \\
\hline\hline
$\alpha$-U&[Xe]4f&6s,6p,5d&12/8&3&11.824  \\
HgSe&Hg: [Kr]4d&Hg: 5s,5p,5d,4f&10/6&Hg: 3&6.946  \\
& Se: [Ne]&Se: 3s,3p,3d&&Se: 2&  \\
HgTe&Hg: [Kr]4d&Hg: 5s,5p,5d,4f&10/6&Hg: 3&6.946  \\
& Te: [Ar]3d&Te: 4s,4p,4d&&Te: 2&  \\
FePt&Fe: [Ne]&Fe: 3s,3p&10/8&Fe: 3&12.0  \\
& Pt: [Kr]4d&Pt: 5s,5p,4f&&Pt: 3&  \\
Th&[Xe]&5s,6s,5p,6p,5d,4f&10/10&3&12.0  \\
\end{tabular}
\end{center}
\end{table}

\begin{figure}[t]
\centering
\includegraphics[width=9.0 cm]{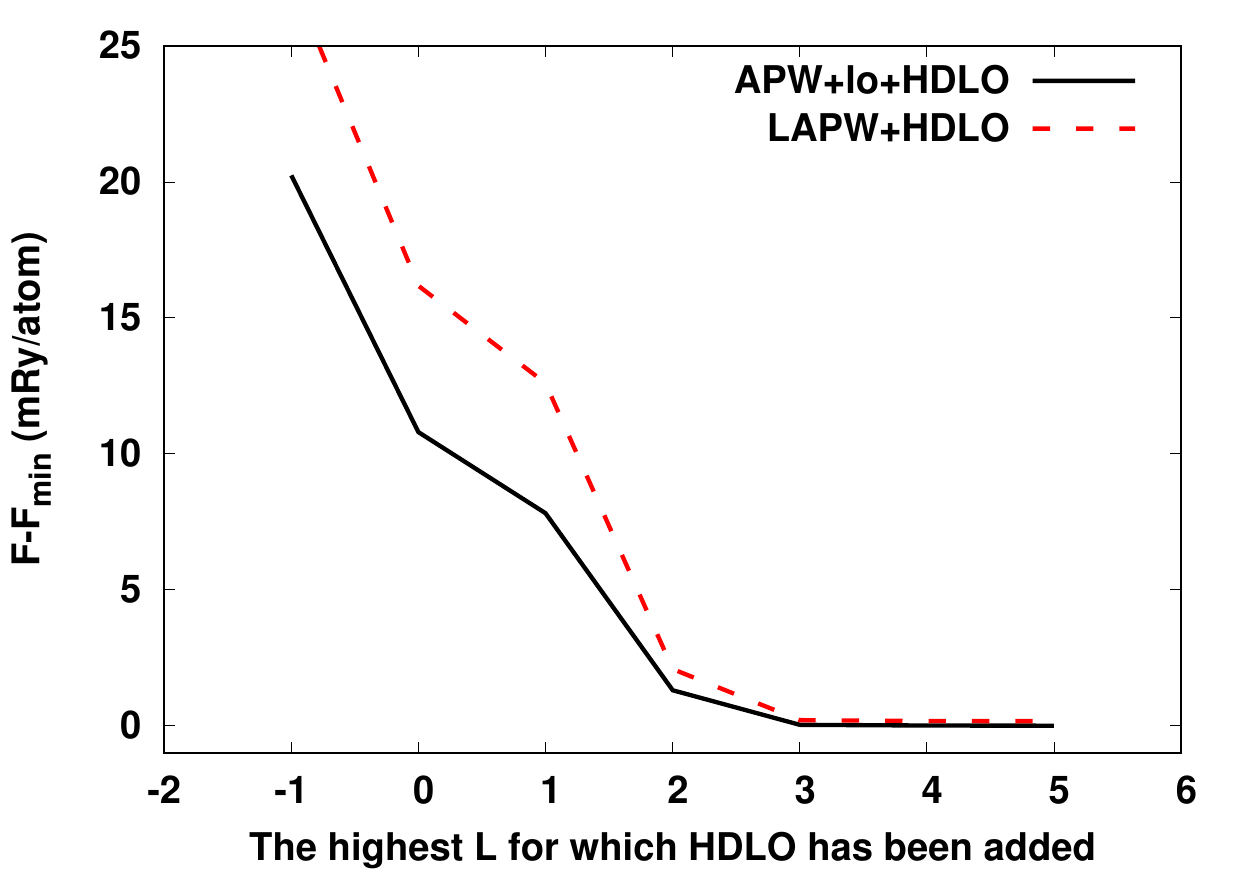}
\caption{Effect of adding the HDLO orbitals to LAPW/APW+lo basis set for $\alpha$-U. HDLO orbitals were added subsequently for L=0, 1, 2, 3, 4, and 5. Value L=-1 at X-axis means there were no HDLO added. The constant $F_{min}$=-112276.465428 Ry was used for all lines.} \label{l_hd}
\end{figure}

\begin{figure}[t]
\centering
\includegraphics[width=9.0 cm]{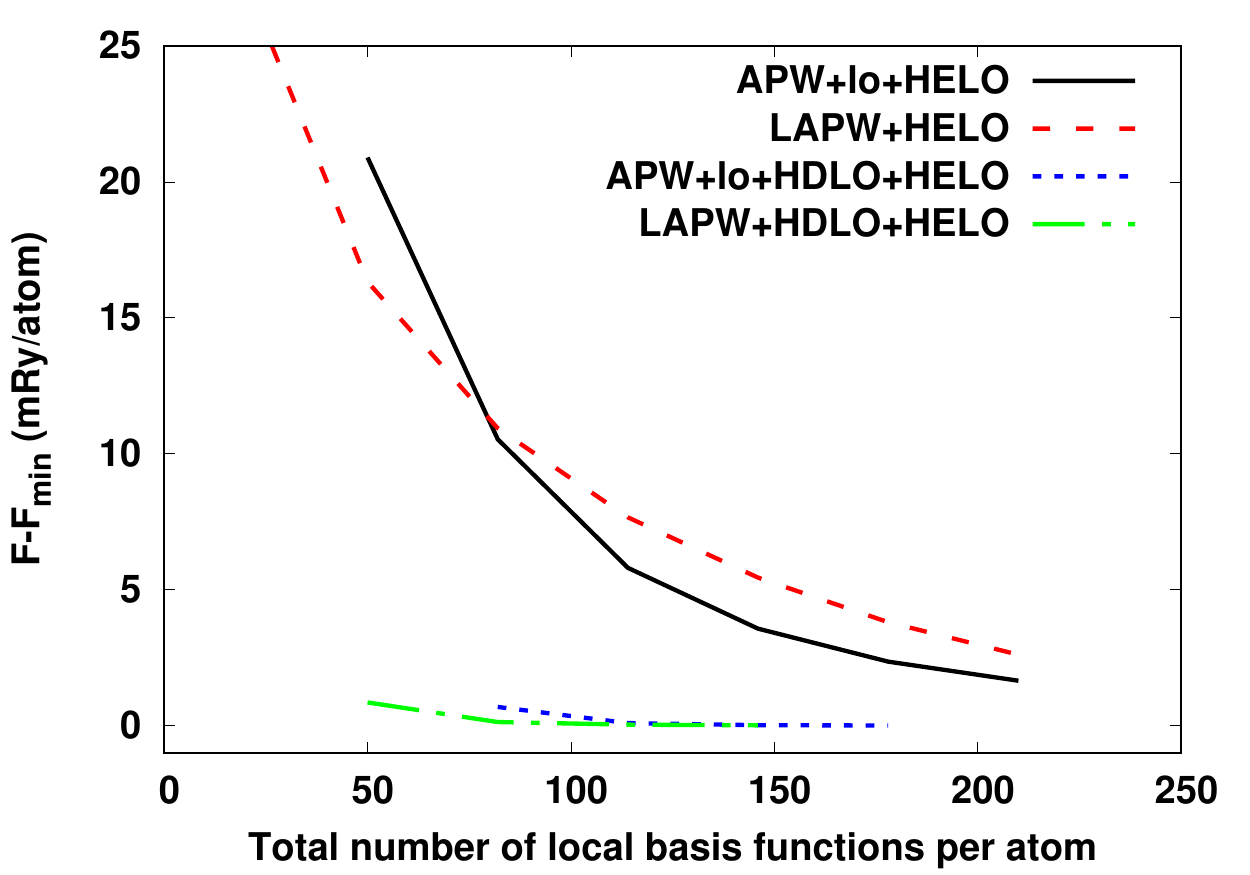}
\caption{Effect of adding the high energy HELO orbitals to different basis sets for $\alpha$-U. HELO orbitals were added by groups. Each group was included by a simultaneously adding HELO's to the basis set for L=0, 1, 2, and 3 (i.e. 32 functions per atom). Total number of local orbitals (X axis) includes semicore 6s, 6p, and 5d states. Different starting values of the curves correspond to the fact that LAPW basis (without high energy HELO's) has only semicore states, whereas APW+lo has, in addition, 32 'lo' orbitals, LAPW+HDLO has, in addition, 32 HDLO's, and APW+lo+HDLO has, in addition, 64 'lo,HDLO' functions. The constant $F_{min}$=-112276.466738 Ry was used for all lines.} \label{he_add}
\end{figure}

\begin{figure}[b!]
\centering
\includegraphics[width=9.0 cm]{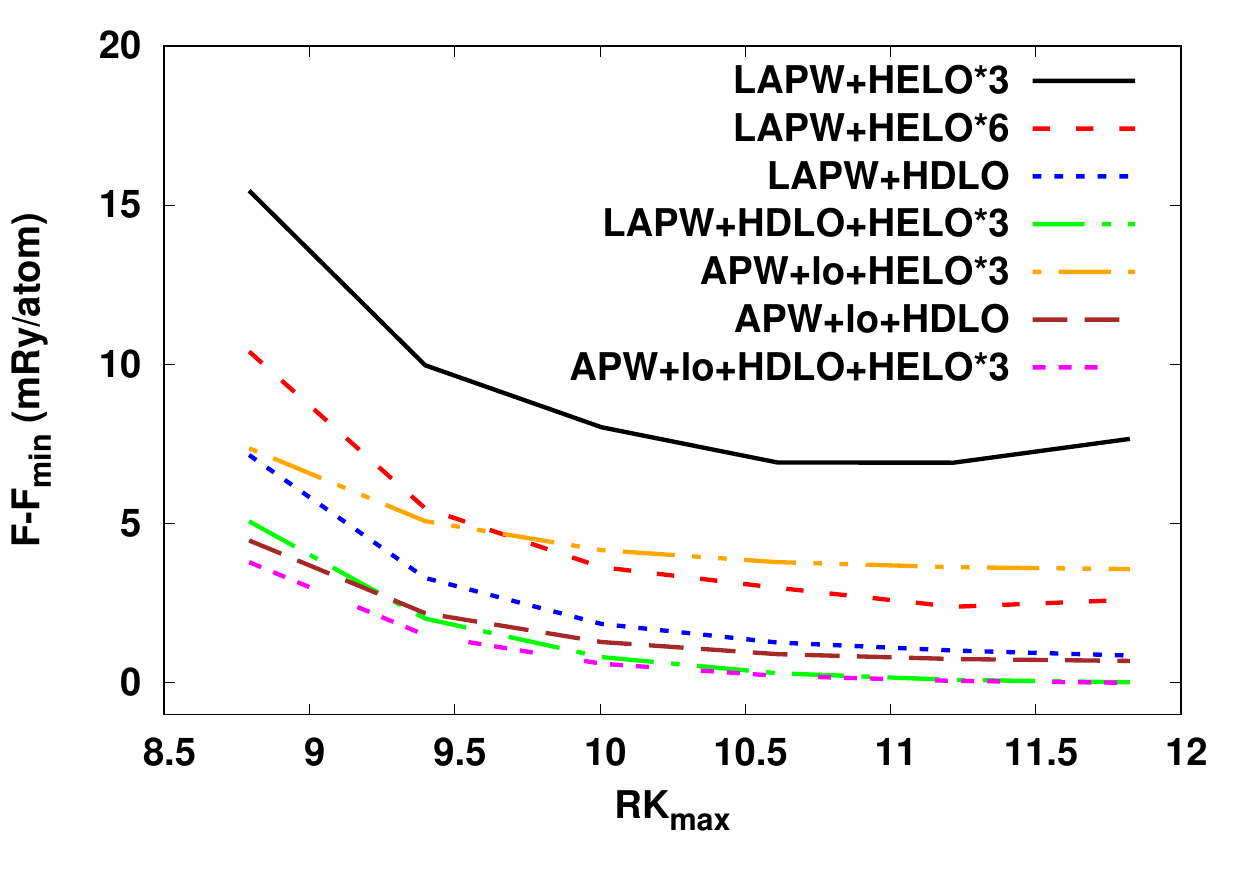}
\caption{Dependence of the electronic free energy on the parameter RK$_{max}$ which defines basis set in the interstitial region. $\alpha$-U case. The constant $F_{min}$=-112276.466738 Ry was used for all lines.} \label{pw_dep}
\end{figure}

This section presents results of the calculations. In order to make presentation more compact, principal structural parameters for studied solids have been collected in Table \ref{list_s} and most important set up parameters have been collected in Table \ref{setup_s}. All calculations have been performed with $T=300K$. Whenever the simplified approach for relativistic effects appears below in tables or figures, it will be marked with (SRA). The Local Density Approximation (LDA) as parametrized by Perdew and Wang\cite{prb_45_13244} was used in all calculations (if not specified otherwise).

Essential part of this work is dedicated to checking of the performance of LAPW/APW+lo basis sets supplemented with different kind of local orbitals (HDLO and/or HELO) as implemented in fully relativistic branch of the FlapwMBPT code. In some of the Figures, the abbreviation HELO comes together with a factor, like HELO*n. It simply means that n sets of HELO's was added to each orbital channel which exists in a free atom. For instance, for actinides it would be s, p, d, and f channels. Most of the calculations have been performed for $\alpha$-uranium which combines strong relativistic effects and sufficiently large interstitial region (the biggest fraction of the MT volume is 52.7\% which one can compare to the fracture of 74\% in the face centered cubic structure). First circumstance makes it important to use an approach based on Dirac's equation. Second circumstance makes it important to use basis sets with flexible representation of the interstitial region, such as APW/LAPW.

Let us begin exploring which orbital channels (s, p,d, f, or all of them) are needed to be supplemented with HDLO orbitals. Remarkable improvements in accuracy discovered by authors of Ref. \cite{cpc_184_2670} and by the authors of Ref. \cite{cpc_220_230} with addition of HDLO's makes it important to consider this extension every time when one wishes to reach well converged results. However, conclusions on how exactly HDLO's have to be added are somewhat different in Ref. \cite{cpc_184_2670} and in Ref. \cite{cpc_220_230}. In the first paper, HDLO's were added to all important channels (spdf) using cerium as an example. In the second paper, authors recommend to use HDLO's only for d or f channel and when the MT radius is sufficiently large (2.5 a.u. or larger). This recommendation is based on their study of a number of materials representing different blocks of the periodic table. In this work, the question was addressed by adding HDLO's in step by step fashion, namely, by adding them first to s-orbitals, then to p orbitals (keeping them in s-orbitals), and so on, till L=5. The results are presented in Fig.\ref{l_hd}. As one can see, the most dramatic effect on the electronic free energy was obtained when HDLO of s-type was added which can be considered as contradiction to the rule established in Ref. \cite{cpc_220_230}. Actually, the effect of addition of f-type HDLO is quite moderate in case of $\alpha$-U. Also, the HDLO's with L$>$3 were of no effect and, correspondingly, were not included in all subsequent calculations.

Next graph (Figure \ref{he_add}) demonstrates the effect of adding the High Energy Local Orbitals (HELO). Semicore states were always included and they are not studied here. As one can easily conclude, addition of HELO's should be done only when a set of HDLO's is already in the basis set. Without HDLO's being already included, both LAPW and APW+lo show very slow convergence when they are being supplemented with increasing number of HELO's. It looks like the authors of Ref. \cite{cpc_184_2670} have dismissed an opportunity to add HDLO's to the APW+lo basis set before adding HELO's. It is true that LAPW+HDLO basis set and APW+lo basis set are of the same size, which might suggest to proceed with adding HELO's to both of them. However, for instance, LAPW+HDLO+HELO*1 and APW+lo+HDLO are also of the same size with the second set being more efficient. Addition of more HELO's eventually makes the difference between LAPW+HDLO+HELO's and APW+lo+HDLO+HELO's disappear, but the rate of convergence is higher with APW+lo as original basis set.

Figure \ref{pw_dep} shows the convergence of the total energy with respect to the parameter $RK_{max}$ which governs the size of the basis set in the interstitial region. In this figure, one can see how an addition of HDLO/HELO to LAPW and APW+lo affects the convergence with respect to the $RK_{max}$. Because of stricter augmentation constraints, LAPW basis set is prone to linear dependencies at large $RK_{max}$ which reveals itself here by a slight upturn of the curves LAPW+HELO*3 and LAPW+HELO*6. Adding more of HELO's reduces upturn but very slowly. This drawback of the LAPW basis set essentially disappears when one includes HDLO's to the basis set. As one can see, the curve LAPW+HDLO (without any HELO's) is below of the curve LAPW+HELO*6 (with 6 sets of HELO's!!!). APW+lo seems to be a better starting point for adding LO's. It doesn't show upturns, i.e. it is a lot less prone to linear dependencies. It is interesting, however, that with HDLO's in the basis, the differences between LAPW and APW+lo gradually disappear (at large $RK_{max}$ and sufficient number of HELO's). Still, the calculations based on APW+lo show faster convergence with respect to $RK_{max}$. As it was said before, this is a result of additional flexibility of APW's as compared to LAPW's related to less restrictive augmentation\cite{prb_64_195134}. Obviously, this conclusion depends on the radius of the MT spheres. For solids with smaller fraction of the MT volume the advantage of APW+lo increases, and vice versa.

\begin{figure*}[h]       
    \fbox{\includegraphics[width=7.5 cm]{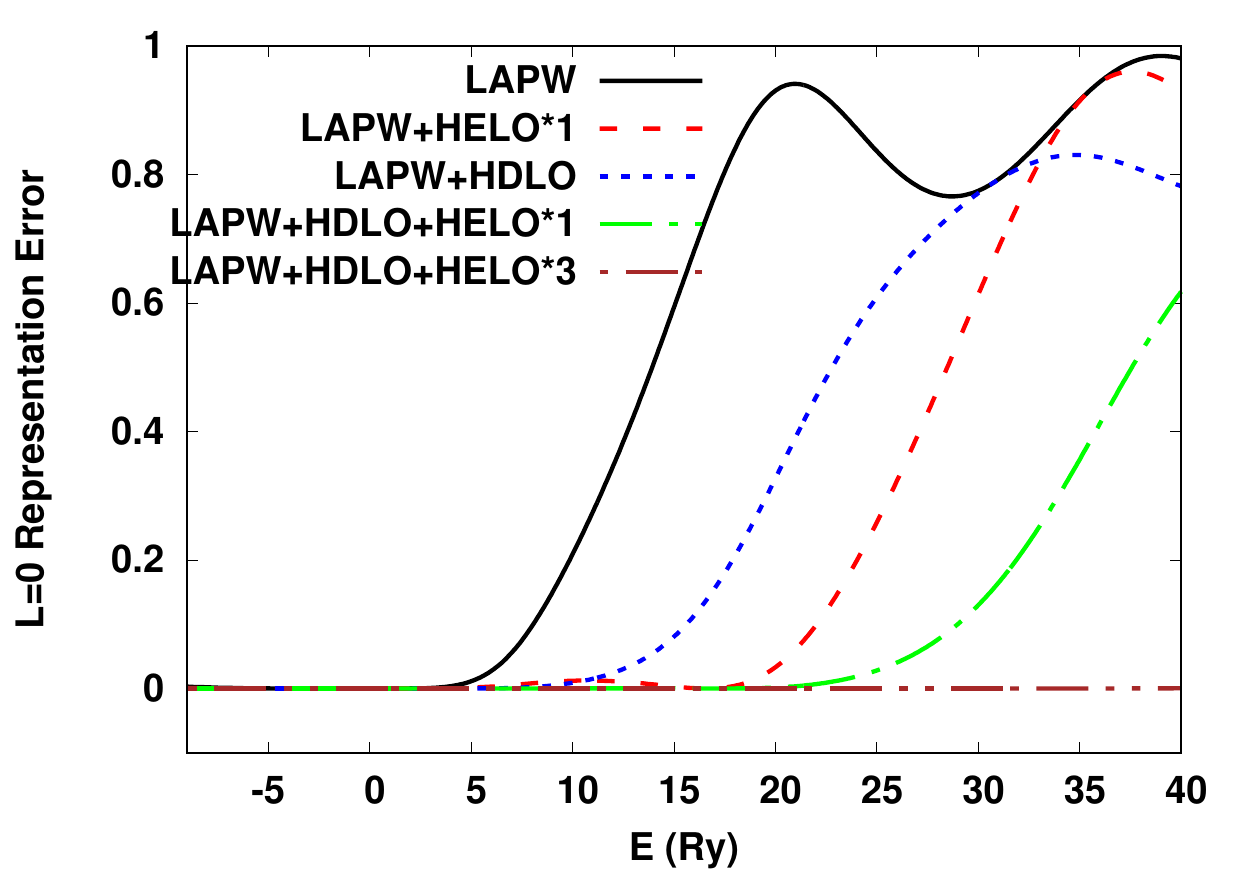}}   
    \hspace{0.02 cm}
    \fbox{\includegraphics[width=7.5 cm]{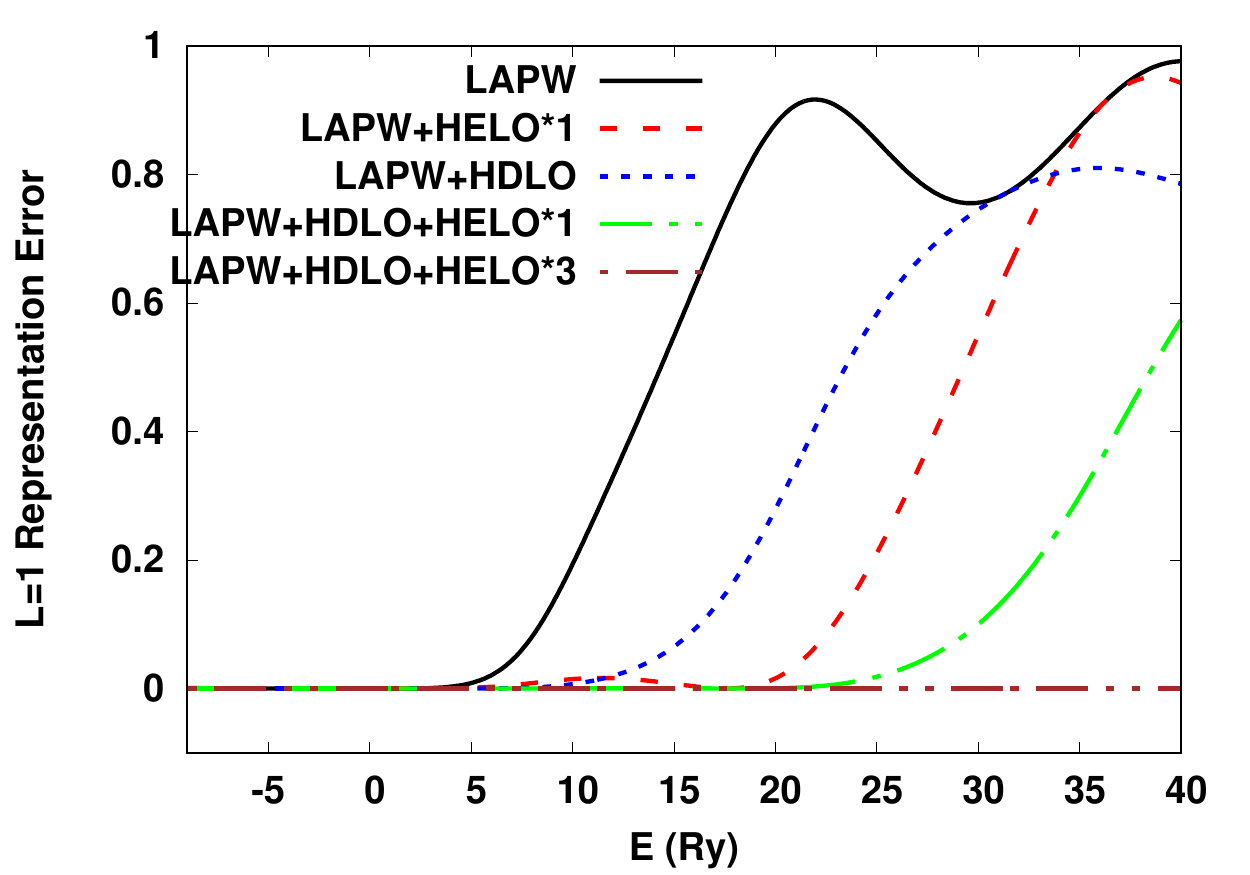}}  
    \hspace{0.02 cm}
    \fbox{\includegraphics[width=7.5 cm]{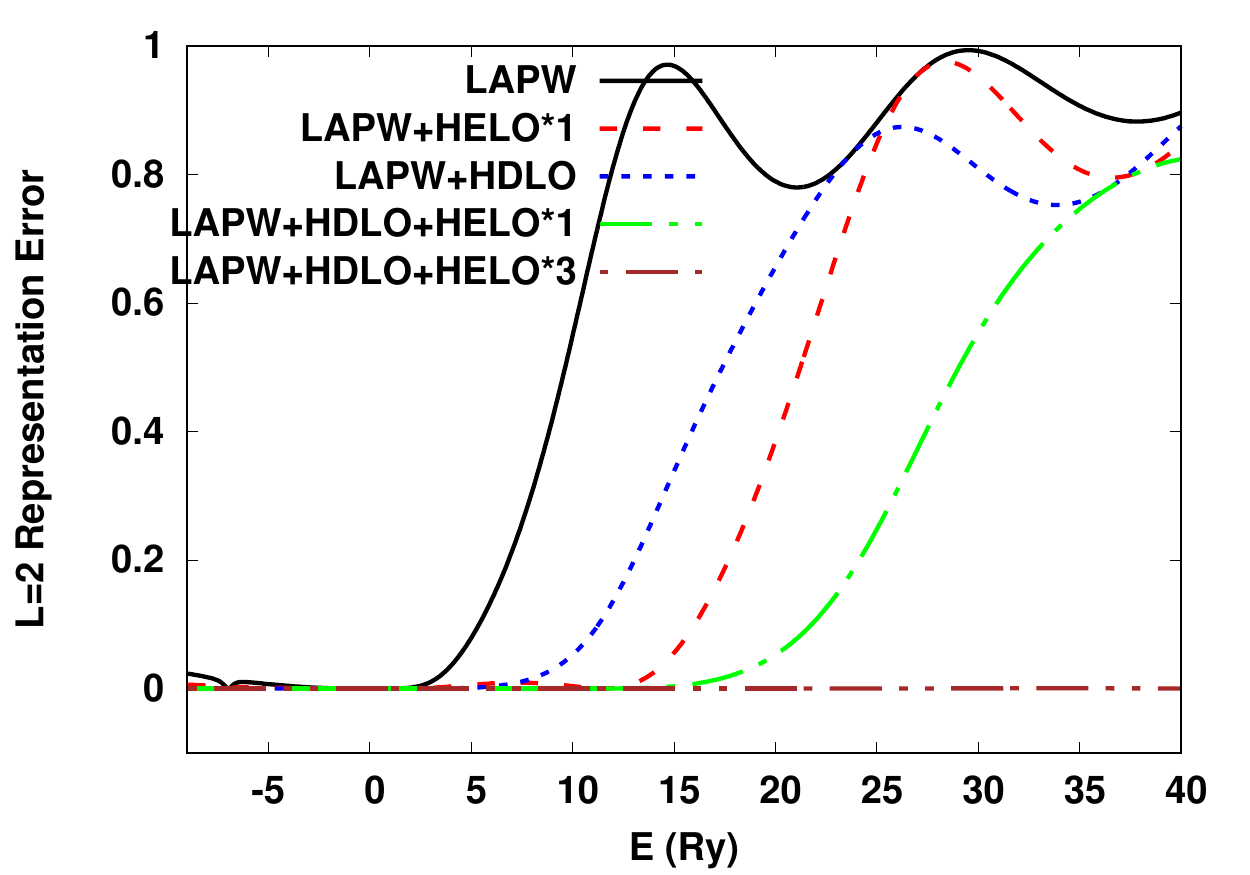}}  
    \hspace{0.02 cm}
    \fbox{\includegraphics[width=7.5 cm]{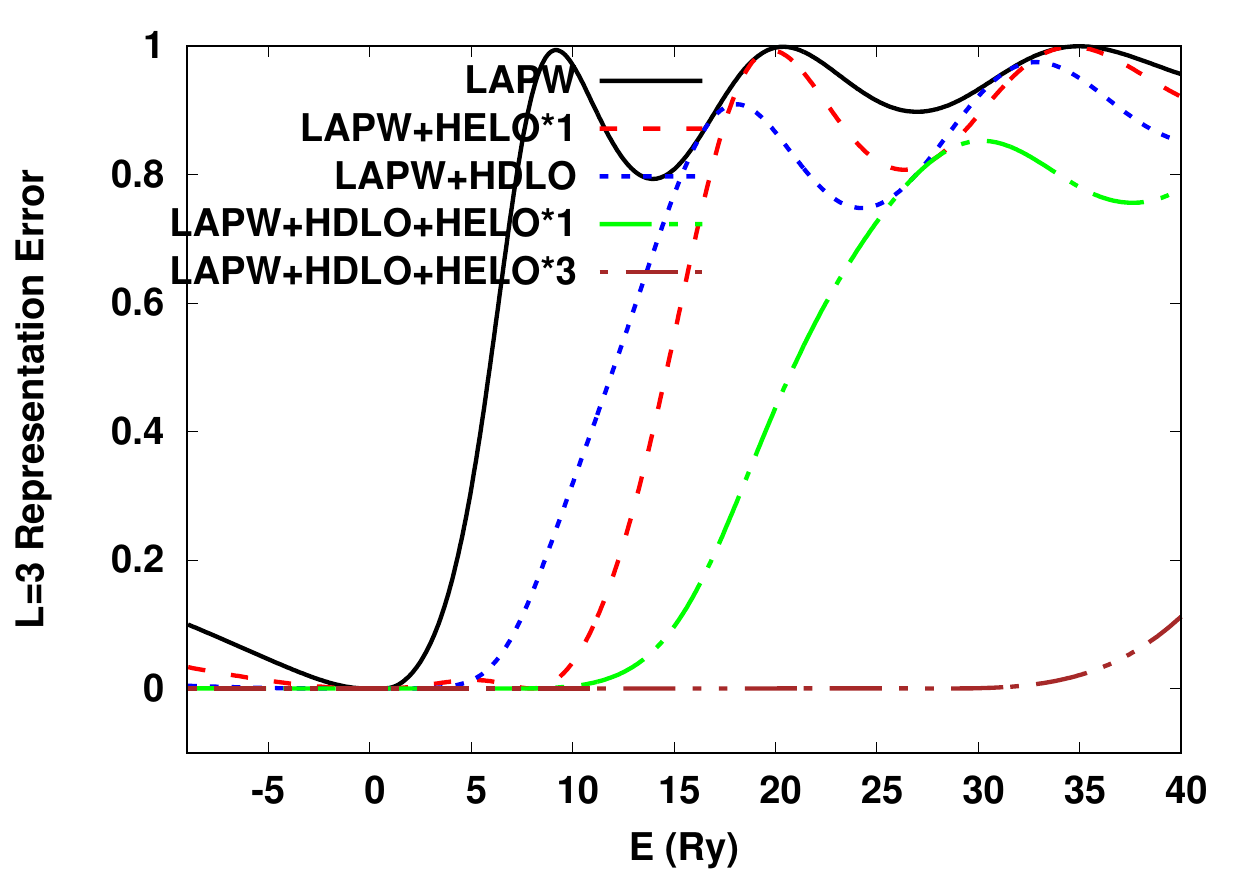}}
    \caption{Representation errors of the basis sets. Top left graph is for s-orbitals, top right is for p-orbitals, bottom left is for d-orbitals, and bottom right is for f-orbitals. The case of $\alpha$-U.}
    \label{repr_e}
\end{figure*}

\begin{figure}[h]       
    \fbox{\includegraphics[width=6.5 cm]{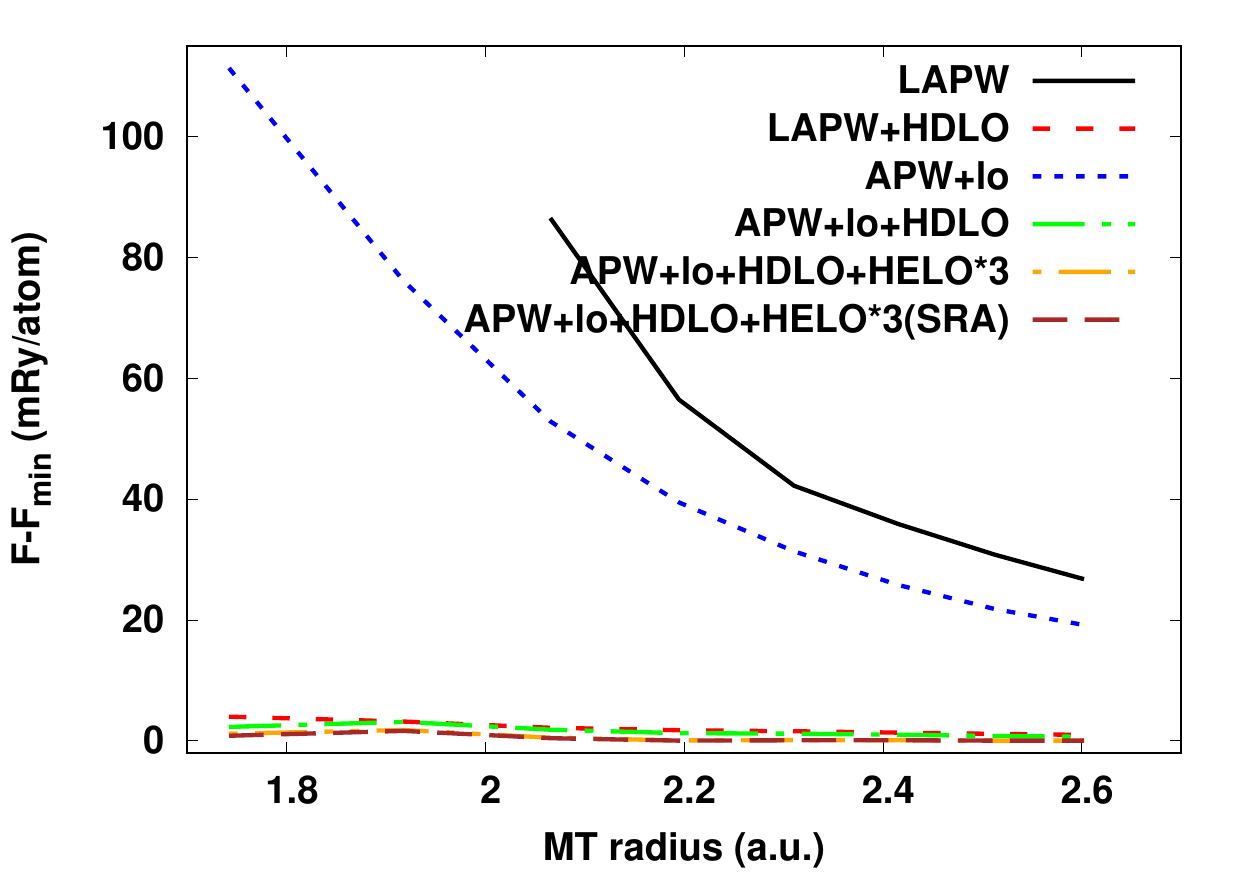}}   
    \hspace{0.02 cm}
    \fbox{\includegraphics[width=6.5 cm]{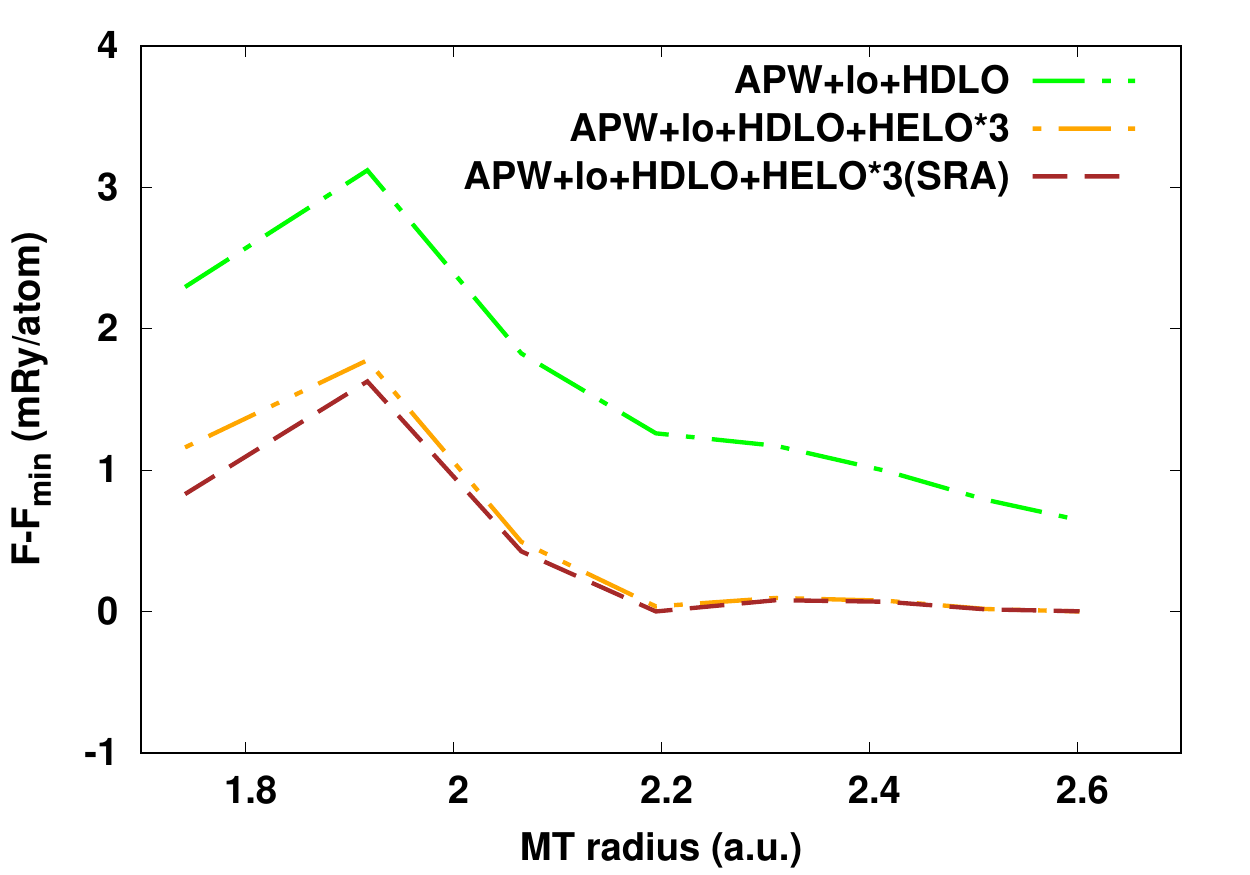}}
    \caption{Dependence of the electronic free energy on the radius of MT sphere for $\alpha$-U. The constant $F_{min}$=-112276.468421 Ry was used for all lines.}
    \label{e_mt_dep}
\end{figure}

In order to shed some light on the reasons of why the addition of HDLO's should precede the addition of HELO's the study of the so called representation error $\Delta_{l}(E)$ which was introduced by authors of Ref. \cite{cpc_184_2670} has been performed. The same definition of $\Delta_{l}(E)$ as in Ref. \cite{cpc_184_2670} is used, namely

\begin{equation}\label{re_er}
\Delta_{l}(E)=\sqrt{\int_{0}^{S_{MT}}dr r^{2}[u_{li}(r,E)-\tilde{u}_{li}(r,E)]},
\end{equation}
where $u_{li}(r,E)$ is a solution of radial equation for a given energy E, whereas $\tilde{u}_{li}(r,E)$
is the best representation of $u_{li}(r,E)$ in terms of the basis functions inside a given MT sphere. Representation error (\ref{re_er}) doesn't take into account the details of the augmentation restriction and, as a result, provides only a lower bound of the error. In particular, it should not be used for comparing LAPW and APW+lo accuracy because augmentation restrictions are different. It is useful, however, to find out about problematic l-channels when one uses a specific, LAPW for instance, augmentation. Figure \ref{repr_e} provides such analysis in the case of LAPW supplemented with different local functions. One can see that the accuracy of the LAPW basis set is poor in the region of semicore energies (d and f orbitals) and at high energies (all orbitals). High energies are not important when one considers DFT free energy, but semicore region is very important. It is obvious from the figure, that addition of HDLO's is better than addition of HELO's, because the HDLO's basically remove the error in the semicore region whereas HELO's only reduce it. Also, in high energy region, addition of the HELO's leaves a hump at about 5Ry whereas addition of the HDLO's consistently pushes up the limit of energies where the basis is complete. It is important to note (for GW applications) that 3 HELO's allow to consider the basis set as complete up to 32 Ry for f-states, and even up to higher energies for spd states.

One more example of the importance of HDLO orbitals provides the study of a dependence of the calculated electronic free energy on the MT radius (Figure \ref{e_mt_dep}). Considering that uranium has a big core, the reducing of the radius by more than 30\% represents very stringent test of the basis set and of the numerical algorithms implemented in the code. For this particular test, parameter $RK_{max}$ was increased to 12.5, and core orbitals 5s and 5p were moved to the semicore list. Basis sets, LAPW and also APW+lo, which normally are considered as quite accurate, demonstrate their insufficiency at small MT radiuses. Remarkably, however, that addition of just one set of HDLO orbitals essentially removes dependence on the choice of radius of the MT sphere. Detailed graph (right hand side) shows that 3 sets of HELO's improve the accuracy of energy only very moderately. Deviation from the smooth behavior seen in this graph is related to two factors. First factor is that the rule "constant $RK_{max}$ should be kept when radius changes" begins to break at very small MT radii. At very small radii one should take more plane waves than this rule dictates. Second factor is that 4f core level becomes not exactly confined inside MT sphere and should be included in the semicore states in very precise calculations. For our present discussion, however, the level of accuracy presented in Fig. \ref{e_mt_dep} is sufficient.

The above discussion shows that the modern extensions of the APW+lo/LAPW basis set, such as HDLO and HELO, allow to reach very high accuracy. The fact that they allow to essentially eliminate the representation (linearization) error (Fig. \ref{repr_e}) means that the dependence on the exact positions of the traditional LAPW linearization centers (energies which we use to solve the radial equations) can be made to be very weak or to be totally negligible. However, the computer code still has to decide where to place them. Plus, there might be situations when user, by some reason, wants to use a "good old" variant of the basis set without modern extensions. In this case, it is important to place the centers carefully. Also, irrespectively of the accuracy level, one has to be careful in order to avoid linear dependence between the orbitals corresponding to different linearization centers but to the same orbital momentum channel. In this respect, the FlapwMBPT code, as many other codes, has the default option. In the FlapwMBPT code, the linear dependence is avoided by using one linearization center for each principal quantum number (for each orbital momentum L). This is easily achieved by the requirement that the logarithmic derivative of the solution (of the radial equation) should be the same for all principal quantum numbers included (but the number of zeros in the solution is different, of course). By default, the value -L-1 is used for the logarithmic derivative D, but this can be changed by a user. When only one linearization center is used (for a given L) the results can still be dependent on the specific choice of D, even with APW+lo+HDLO/LAPW+HDLO. But the dependence is much smaller than in original LAPW basis set. When one adds a few high energy LO's (HELO's), the dependence disappears. The reason is clearly seen from the Fig. \ref{repr_e}.

Figure \ref{e_mt_dep} also presents the results obtained with simplified treatment of the relativistic effects. As one can see, the effect of simplification is extremely small, and can be considered as negligible in this case. In order to show that this conclusion might be quite general, three more materials were included in the study. For mercury selenide (HgSe) and mercury telluride (HgTe) the effect of spin-orbit splitting was studied (Table \ref{Hg_res}). Theoretical calculations at DFT level underestimate the spin-orbit splitting (especially in case of HgSe) which suggests importance of many-body effects not included at this level of theory. For our study, however, it is important that the results obtained are consistent with other theoretical studies, particularly with the results obtained in Ref. \cite{prm_1_033803} using SR+SOC+$p_{1/2}$ approach which is the closest (in terms of inclusion of the relativistic effects) to the Dirac equation based approach used in the present work. Even more important is the fact, that again, simplified treatment of the relativistic effects make practically no difference as compared to the fully relativistic calculations.

The next test of the applicability of the simplified approach for relativistic effects was conducted for the case of magnetic crystalline anisotropy effect in FePt. This particular property of FePt was studied many times both theoretically \cite{prb_52_13419,prb_63_144409,prb_94_144436} and experimentally \cite{prb_63_144409,prb_94_144436}. Theoretical results obtained at DFT level consistently overestimate the experimental data (2.258-3.2 meV versus 0.88-1.3 meV) which, again, suggests on the importance of many-body effects beyond DFT. However, as Table \ref{FePt_conv} shows, the results obtained in this work are consistent with previous studies. Also, as in the case of $\alpha$-U and HgSe/HgTe, simplified (SRA) inclusion of relativistic effects works remarkably well here, closely following the fully relativistic approach when the number of k-points in the Brillouin zone changes.

\begin{table}[t]
\begin{center}
\caption{Comparison of the calculated spin-orbit splitting ($\Delta$) and inverse direct band gap ($E_{g}$) at $\Gamma$ point for HgSe and Hgte. All results are in eV. LDA and GGA results are separated by slash (LDA/GGA). If only LDA (GGA) result is available, it is given on the left (right) side of the slash correspondingly. FRA and SRA results differ only by $1\div 2$ meV, so they are identical with the number of digits presented in this table.} \label{Hg_res}
\small
\begin{tabular}{@{}c c c c c}  &\multicolumn{2}{c}{HgSe} & \multicolumn{2}{c}{HgTe}\\
  &$E_{g}$ & $\Delta$  &$E_{g}$ & $\Delta$\\
\hline\hline
SR+SOC\cite{prb_71_045207} &-1.23/ &0.23/  &-1.17/ & 0.80/  \\
SR+SOC\cite{prb_84_205205} &-1.18/ &0.24/  &-1.20/ & 0.78/  \\
SR+SOC\cite{prb_84_085144} &-1.27/ &0.23/  &-1.20/ & 0.78/  \\
SR+SOC\cite{prm_1_033803} &  & /0.22 &  &  /0.71  \\
SR+SOC+$p$\cite{prm_1_033803} & & /0.23  & & /0.78  \\
Dirac, & & & &  \\
this work &-1.27/-1.03 &0.24/0.22 &-1.19/-0.98 & 0.80/0.77  \\
Experiment\cite{prb_61_R5058}   && &-0.32 & 0.91\\
Experiment\cite{ssc_34_441}   &-0.273 & 0.39$\pm$0.01  &&\\
\end{tabular}
\end{center}
\end{table}

\begin{table}[b]
\caption{Magneto-crystalline anisotropy energy (meV) of FePt calculated with fully relativistic approach and with approximate account of the relativistic effects where they were neglected in the interstitial region. Dependence of the results on the total number of points in the Brillouin zone is given.} \label{FePt_conv}
\begin{center}
\begin{tabular}{@{}c c c c c c c} Number of k-points &384 & 800  &1296 & 2156&3072 & 6000 \\
\hline\hline
Dirac eq-n in full unit cell&2.71 &2.88 &3.35 &3.10  &3.24 &2.98  \\
Dirac eq-n in MT &2.69 &2.87 & 3.35 &3.10  &3.24 &2.98 \\
\end{tabular}
\end{center}
\end{table}

\begin{figure*}[h]       
    \fbox{\includegraphics[width=7.5 cm]{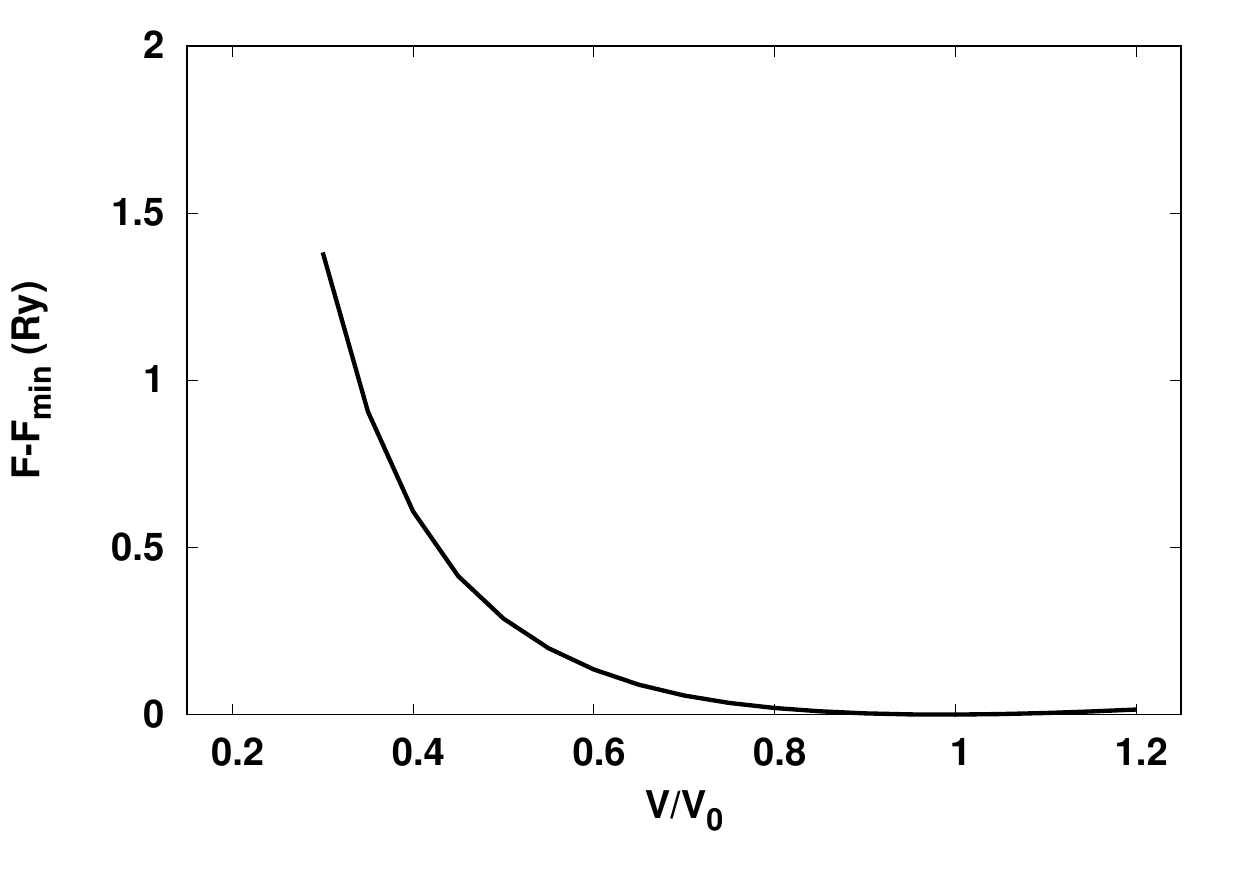}}   
    \hspace{0.02 cm}
    \fbox{\includegraphics[width=7.5 cm]{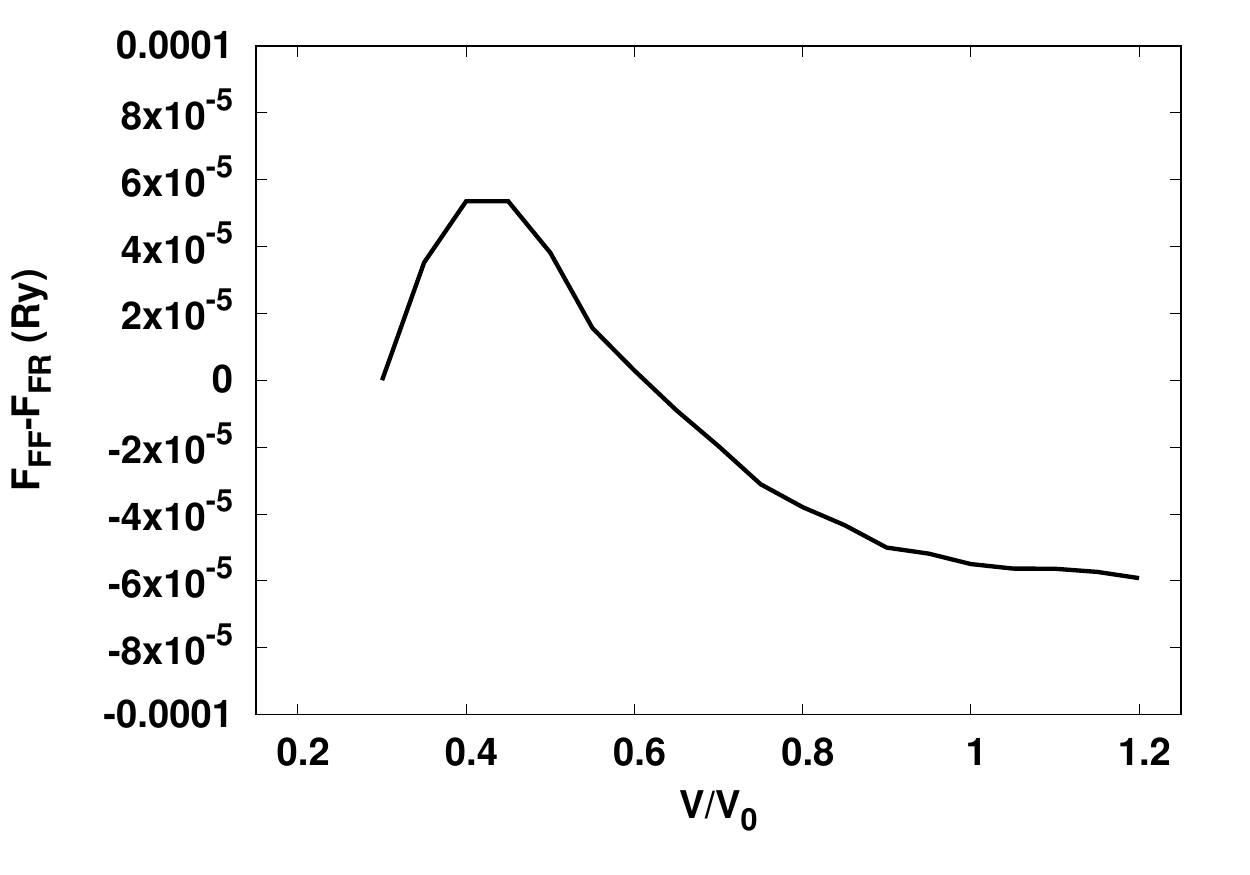}}  
    \hspace{0.02 cm}
    \fbox{\includegraphics[width=7.5 cm]{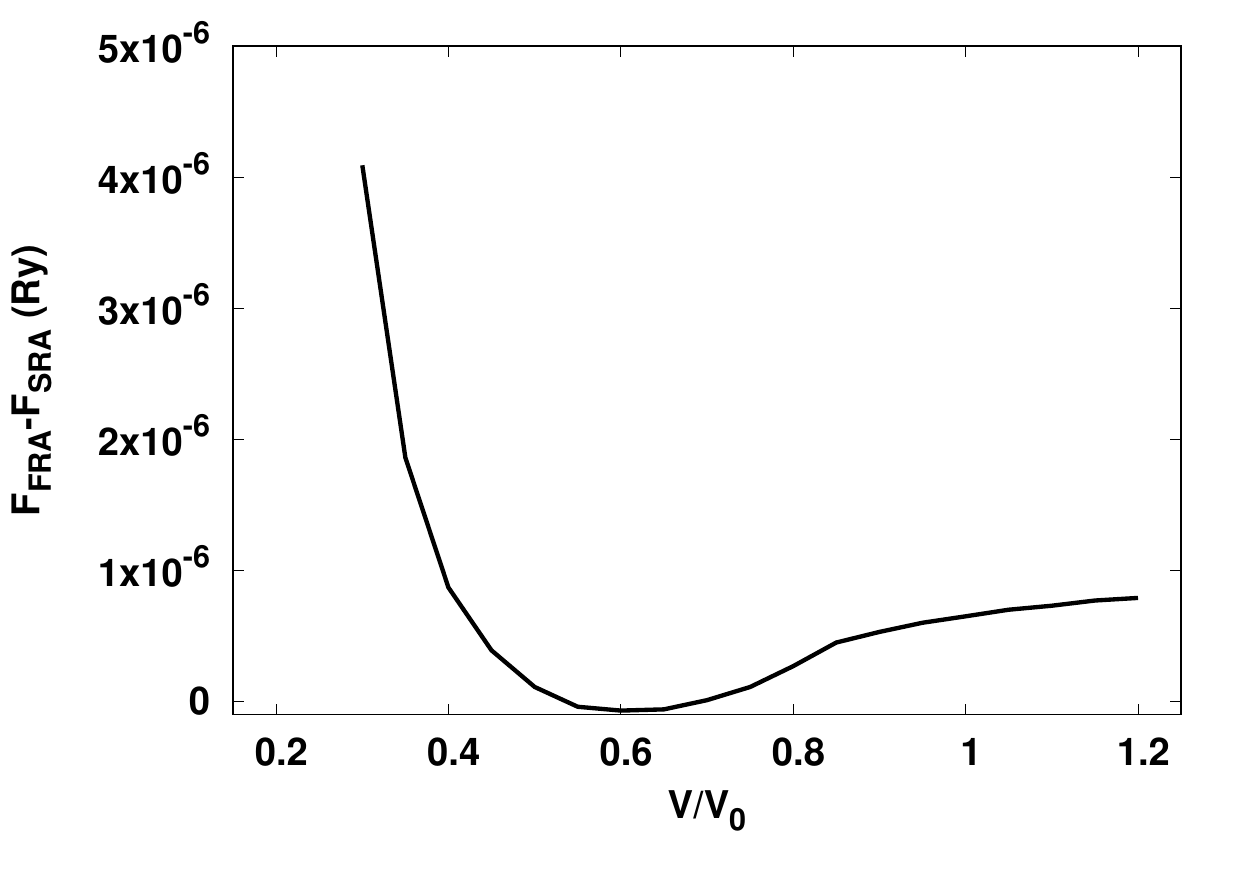}}  
    \hspace{0.02 cm}
    \fbox{\includegraphics[width=7.5 cm]{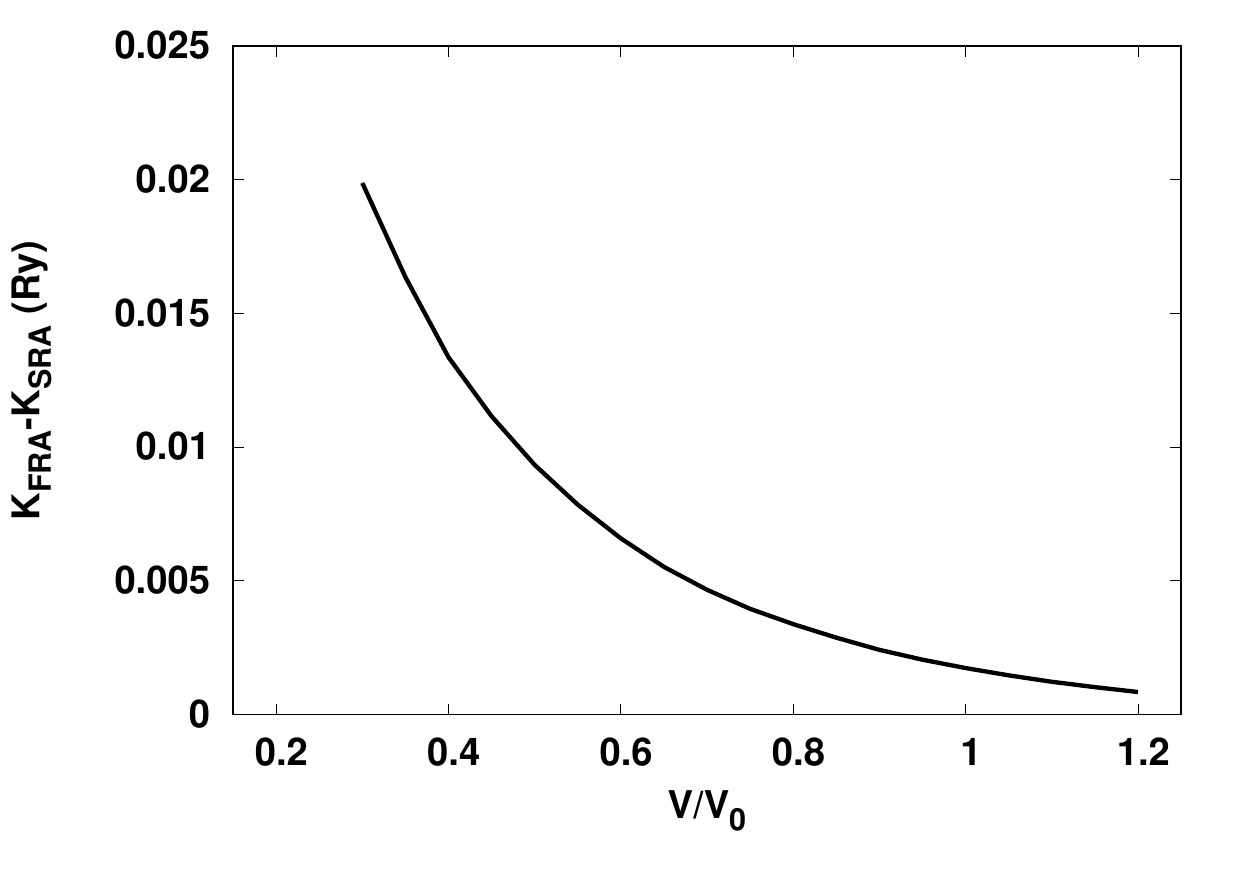}}
    \caption{Results for Thorium. Top left: electronic free energy versus relative volume (V/V$_{0}$). Constant value -53073.499953 Ry has been subtracted from all energies. Top right: Difference of the electronic free energy obtained in FRA with fixed fraction (FF) and fixed MT radius (FR) versus V/V$_{0}$. Bottom left: Difference of the electronic free energy obtained in FRA and in SRA versus V/V$_{0}$. Bottom right: Difference of the electronic kinetic energy (K) obtained in FRA and in SRA versus V/V$_{0}$.}
    \label{Th_res}
\end{figure*}

Motivated by rather small deviations between the results obtained in FRA and in SRA for $\alpha$-U, HgSe, HgTe, and FePt, a deeper analysis of these two approaches (FRA and SRA) was conducted using actinide metal Th as an example. Thorium was selected because it is usually used as a test bed for an assessment of the quality of approximations for the spin-orbit interaction\cite{prb_63_035103,prb_64_153102,jcm_15_2607,prb_101_085114} and also because it has simple crystal structure. Anticipating small differences between FRA and SRA at volumes close to the equilibrium one, the calculations were performed up to very high compression (the lowest V/V$_{0}$ considered was 0.3). As variants of the calculation, we included not only FRA and SRA, but also the calculations performed with fixed fraction (FF) of the MT volume (touching MT spheres) and with fixed MT radius (FR) which was the radius of the touching MT spheres at V/V$_{0}=0.3$. To be in line with other works which study Th, the Generalized Gradient Approximation (GGA) as parametrized in the Ref. [\onlinecite{prl_77_3865}] was used as the exchange correlation potential for this analysis. The Brillouin zone was sampled with $9\times 9\times 9$ mesh of \textbf{k}-points. Other setup parameters are summarized in Tables \ref{list_s} and \ref{setup_s}. In Figure \ref{Th_res} we present the results of the analysis. First graph (top left) shows electronic free energy as a function of the relative volume. On the scale of this curve, the differences between the variants of calculation are absolutely negligible. This curve is included mostly for the purpose of reference and for giving an idea of the scale of energy differences in our analysis. Second graph (top right) represents the difference in the electronic free energy between FF and FR variants of the calculation. It shows one more time the high level of the accuracy which one can attain with the FlapwMBPT code. Even with very large differences between the MT radii in the FF and in FR variants at extended volumes, the free energy difference remains less than $10^{-4}$ Ry. It is worth to point out that in similar tests performed with FPLMTO code\cite{prb_101_085114}, the fixed MT radius was selected at V/V$_{0}=0.82$. Correspondingly, in FPLMTO calculations, the differences between the MT radii in the FF and in FR variants were relatively small. The energy differences, however, were larger than in our case as one can judge from the Figure 2 in the Ref. [\onlinecite{prb_101_085114}]. Two more graphs, presented in Figure \ref{Th_res}, show the most interesting information for our analysis of the SRA. In the bottom left graph, one can see the difference between the electronic free energy obtained in the FRA and in the SRA. In the bottom right graph, one can see the difference between the electronic kinetic energy, obtained in the same approximations. The reason for presenting the kinetic energy is twofold: i) it is well known that total (free) energy, being a variational quantity, is very stable to the variations in a basis set, numerical algorithms etc., whereas its components (such as kinetic energy) are not variational and, as a result, they fluctuate a lot stronger when one changes the calculation setup; ii) kinetic energy is, actually, a direct measure of the relativistic effects as one can see from the Eq. \ref{hkin}. Looking at the graph with the electronic free energy, one can appreciate how little the FRA and the SRA results differ. Their difference is less than $10^{-6}$ Ry for most volumes. Only when the relative volume becomes less than 0.4 the difference starts to increase indicating the fact that we finally are hitting the area near nucleus where relativistic effects are strong. As one can see from the graph which shows the kinetic energy, the FRA and the SRA results differ a lot more than in the case of the free energy, in average by a factor of 10000. Thus, from this comparison, one can draw the following conclusion. The effect of neglecting the relativistic effects in the interstitial region, measured directly by the difference in the kinetic energy, is noticeable (0.001-0.01 Ry, depending on the volume). However, in all practical applications, we are interested, in fact, in its effect not on the kinetic but rather on the total (free) energy which is a variational quantity. So, from the practical point of view, the effect is rather small, and it can be considered almost negligible up to a very high pressure.

The analysis of the relativistic effects in the interstitial region conducted above reveals noticeable contradictions with conclusions of the authors of the Ref. [\onlinecite{prb_101_085114}]. This contradiction has to be discussed. The reason, why authors of the Ref. [\onlinecite{prb_101_085114}] study the subject is that they noticed the problem with 6p states of Thorium in their SR+SOC calculations. The problem, however has already been known for a very long time, since the work by Nordstr\"{o}m et al. \cite{prb_63_035103}. In essence, the problem appears because the SR 6p basis functions have qualitatively different behavior near nucleus as compared to the relativistic (from the Dirac equation) 6p$_{1/2}$ functions. Correspondingly, when these SR 6p states are used as a basis for the estimation of SOC, the result appears to be unstable and sensitive to the boundary conditions at the MT sphere. That is why the change in the MT radius triggers the change basically in everything in SR+SOC calculations: total energy, bands, density of states etc. Beautiful solution (so might be not entirely perfect) of the problem was found by authors of the work [\onlinecite{prb_64_153102}]. They show that the problem disappears if one adds Dirac's 6p$_{1/2}$ functions in the basis set keeping the rest of the basis set in SR form. This simple solution actually tells us that the problem has nothing to do with the interstitial region. Rather, the problem is located quite near the nucleus where SR 6p and the Dirac's 6p$_{1/2}$ states differ most. Later, this conclusion was confirmed in the work [\onlinecite{jcm_15_2607}] using the Dirac equation. Thus, the authors of the Ref. [\onlinecite{prb_101_085114}] draw incorrect conclusions when they make the statements (at least three times in their Conclusion section) like the following: "We also find that the SR + SO $6p_{1/2}$ energy levels depend strongly on the muffin-tin radius, due to the neglect of the SO term in the interstitial." I believe, the above reminder about older works resolves the contradictions between the present work and the work [\onlinecite{prb_101_085114}]. Also, the analysis conducted in the present work serves as a simple and rather direct confirmation of the validity of the older works\cite{prb_64_153102,jcm_15_2607} on the subject of 6p states in Thorium.

\section*{Conclusions}
\label{concl}
Presented work gives detailed account of the implementation of RDFT using basis sets of APW/LAPW type with flexible extensions provided by local orbitals. It is shown that addition of High Derivative Local Orbitals (HDLO's) is extremely efficient in enhancing the accuracy of DFT calculations for $\alpha$-U , as it was earlier discovered for other materials \cite{cpc_184_2670,cpc_220_230}. High Energy Local Orbitals (HELO's), however indispensable for GW calculations, are considerably less efficient in enhancing the accuracy of DFT applications. It was confirmed, using five materials as examples, that simplified handling of the relativistic effects (SRA), namely, considering them only inside the MT spheres, represents a very good approximation. Deeper analysis of the SRA conducted for the electronic free energy of Thorium revealed that high quality of the SRA for this quantity is rooted in the variational property of the free energy.

\section*{Acknowledgments}
\label{acknow}
This work was   supported by the U.S. Department of energy, Office of Science, Basic
Energy Sciences as a part of the Computational Materials Science Program.

\appendix

\section{Variational Schlosser-Marcus principle and its extensions}\label{sm_var}

In LAPW/APW family of methods, basis functions generally have discontinuities. In this case, usual expression for one-electron energies E and wave functions $\Psi$:

\begin{equation}\label{usual_var}
E \int d\mathbf{r} \Psi^{*}(\mathbf{r})\Psi(\mathbf{r})=\int d\mathbf{r} \Psi^{*}(\mathbf{r})H\Psi(\mathbf{r},
\end{equation}
is not variational any more. In order to make it variational, one has to add surface terms. Specific form of the correction surface terms depends on the explicit expression for the kinetic energy operator. In non-relativistic case the variational expression was derived by Schlosser and Marcus \cite{pr_131_2529}:

\begin{align}\label{sm_nr}
E&\int d\mathbf{r} \Psi^{*}(\mathbf{r})\Psi(\mathbf{r})=\int d\mathbf{r} \Psi^{*}(\mathbf{r})[-\nabla^{2}+V(\mathbf{r})]\Psi(\mathbf{r})
\nonumber\\& 
-\frac{1}{2}\sum_{t}\int_{S_{t}}d\mathbf{S}\{[\Psi^{*}_{I}(\mathbf{r})+\Psi^{*}_{t}(\mathbf{r})]
[\nabla\Psi_{I}(\mathbf{r})-\nabla\Psi_{t}(\mathbf{r})]\nonumber\\ &-[\nabla\Psi^{*}_{I}(\mathbf{r})+\nabla\Psi^{*}_{t}(\mathbf{r})]
[\Psi_{I}(\mathbf{r})-\Psi_{t}(\mathbf{r})]\},
\end{align}
where sum runs over all MT surfaces of a solid. Subscripts $I$ and $t$ are used to distinguish, correspondingly, the interstitial and MT representations of the wave functions. Normal vector is directed outside the MT spheres. In the FlapwMBPT code a slightly modified form of (\ref{sm_nr}) is used. The modification is obtained with use of the second Green's Identity and reads as the following:

\begin{align}\label{sm_nr1}
E &\int d\mathbf{r} \Psi^{*}(\mathbf{r})\Psi(\mathbf{r})=\int d\mathbf{r}\Psi^{*}(\mathbf{r})V(\mathbf{r})\Psi(\mathbf{r})\nonumber\\&+\frac{1}{2}\int d\mathbf{r}\{\Psi^{*}(\mathbf{r})[-\nabla^{2}]\Psi(\mathbf{r})+\Psi(\mathbf{r})[-\nabla^{2}]\Psi^{*}(\mathbf{r})\}
\nonumber\\  
&-\frac{1}{2}\sum_{t}\int_{S_{t}}d\mathbf{S}\{\Psi^{*}_{t}(\mathbf{r})\nabla\Psi_{I}(\mathbf{r})-
\nabla\Psi^{*}_{t}(\mathbf{r})\Psi_{I}(\mathbf{r})\nonumber\\&-\Psi^{*}_{I}(\mathbf{r})\nabla\Psi_{t}(\mathbf{r})+\nabla\Psi^{*}_{I}(\mathbf{r})\Psi_{t}(\mathbf{r})\}.
\end{align}

This expression is used in non-relativistic and scalar-relativistic branches of the FlapwMBPT code. We will call this expression as Variational Schlosser-Marcus (VSM) expression. It is also used for SRA implementation with two differences: i) in the volume integral, the relativistic expression for kinetic energy is used inside MT, and ii) in the surface terms the values and derivatives of the big components are assumed. In (\ref{sm_nr1}) the spherical harmonics expansion of the surface terms is explicitly limited to the highest L ($L_{max}$) in the expansion inside MT. In LAPW case, where functions and the first derivatives are continuous at the MT boundary up to $L_{max}$, the surface terms identically disappear. In other words, in LAPW case, the application of VSM is reduced to using the symmetrized matrix elements of the kinetic energy operator.

In the fully relativistic case, the generalized VSM expression was introduced by Loucks \cite{pr_139_A1333} and reads as the following:
\begin{align}\label{sm_rl}
E&\int d\mathbf{r} \Psi^{\dagger}(\mathbf{r})\Psi(\mathbf{r})\nonumber\\&=\int d\mathbf{r} \Psi^{\dagger}(\mathbf{r})
[-ic\boldsymbol{\alpha}\cdot\nabla +(\beta-I)\frac{c^{2}}{2}+I*V(\mathbf{r})]\Psi(\mathbf{r})
\nonumber\\  
&+\frac{ic}{2}\sum_{t}\int_{S_{t}}d\mathbf{S}[\Psi^{\dagger}_{t}(\mathbf{r})+\Psi^{\dagger}_{I}(\mathbf{r})]
\boldsymbol{\alpha}[\Psi_{t}(\mathbf{r})-\Psi_{I}(\mathbf{r})].
\end{align}

With help of the identity

\begin{align}\label{id_rl}
&\int d\mathbf{r} U^{\dagger}(\mathbf{r})[-ic\boldsymbol{\alpha}\cdot\nabla]T(\mathbf{r})=\nonumber\\&\int d\mathbf{r} \{T^{\dagger}(\mathbf{r})[-ic\boldsymbol{\alpha}\cdot\nabla]U(\mathbf{r})\}^{*}
-ic\int_{S}d\mathbf{S}U^{\dagger}(\mathbf{r})\boldsymbol{\alpha}T(\mathbf{r}),
\end{align}
the expression (\ref{sm_rl}) can be transformed to the following equivalent form:
\begin{align}\label{sm_rl1}
E\int d\mathbf{r} \Psi^{\dagger}(\mathbf{r})\Psi(\mathbf{r})&=\int d\mathbf{r} \Psi^{\dagger}(\mathbf{r})
[(\beta-I)\frac{c^{2}}{2}+I*V(\mathbf{r})]\Psi(\mathbf{r})\nonumber\\&+Re\int d\mathbf{r}\Psi^{\dagger}(\mathbf{r})
[-ic\boldsymbol{\alpha}\cdot\nabla]\Psi(\mathbf{r})
\nonumber\\  
&-\frac{ic}{2}\sum_{t}\int_{S_{t}}d\mathbf{S}\Psi^{\dagger}_{t}(\mathbf{r})
\boldsymbol{\alpha}\Psi_{I}(\mathbf{r})\nonumber\\&
+\frac{ic}{2}\sum_{t}\int_{S_{t}}d\mathbf{S}\Psi^{\dagger}_{t}(\mathbf{r})
\boldsymbol{\alpha}\Psi_{t}(\mathbf{r}).
\end{align}

This form of the relativistic VSM is used in the FlapwMBPT code. It is helpful to give the surface correction term a further consideration. For every term in the expansion of the surface term in spherical spinors we can write

\begin{align}\label{sm_term}
T^{t}_{il\mu}&=-\frac{ic}{2}\int_{S_{t}}dS
\left( 
\begin{array}{c}
g^{t}_{il\mu}(\mathbf{r})\\
\frac{i}{c}f^{t}_{il\mu}(\mathbf{r})
\end{array}
\right)^{\dagger}
\boldsymbol{\alpha}\cdot\mathbf{n}
\left( 
\begin{array}{c}
g^{I}_{il\mu}(\mathbf{r})\\
\frac{i}{c}f^{I}_{il\mu}(\mathbf{r})
\end{array}
\right)\nonumber\\&
+\frac{ic}{2}\int_{S_{t}}dS
\left( 
\begin{array}{c}
g^{I}_{il\mu}(\mathbf{r})\\
\frac{i}{c}f^{I}_{il\mu}(\mathbf{r})
\end{array}
\right)^{\dagger}
\boldsymbol{\alpha}\cdot\mathbf{n}
\left( 
\begin{array}{c}
g^{t}_{il\mu}(\mathbf{r})\\
\frac{i}{c}f^{t}_{il\mu}(\mathbf{r})
\end{array}
\right)\nonumber\\
&=\frac{1}{2}\int_{S_{t}}dS\Big[
g^{^{\dagger}t}_{il\mu}(\mathbf{r})\boldsymbol{\alpha}\cdot\mathbf{n}f^{I}_{il\mu}(\mathbf{r})
-f^{^{\dagger}t}_{il\mu}(\mathbf{r})\boldsymbol{\alpha}\cdot\mathbf{n}g^{I}_{il\mu}(\mathbf{r})\nonumber\\&
-g^{^{\dagger}I}_{il\mu}(\mathbf{r})\boldsymbol{\alpha}\cdot\mathbf{n}f^{t}_{il\mu}(\mathbf{r})
+f^{^{\dagger}I}_{il\mu}(\mathbf{r})\boldsymbol{\alpha}\cdot\mathbf{n}g^{t}_{il\mu}(\mathbf{r})\Big]
\nonumber\\
&=-\frac{S^{2}_{t}}{2}\Big[g^{^{*}t}_{il\mu}(S_{t})f^{I}_{il\mu}(S_{t})-f^{^{*}t}_{il\mu}(S_{t})g^{I}_{il\mu}(S_{t})\nonumber\\&
-g^{^{*}I}_{il\mu}(S_{t})f^{t}_{il\mu}(S_{t})+f^{^{*}I}_{il\mu}(S_{t})g^{t}_{il\mu}(S_{t})\Big].
\end{align}

In this form, the surface contribution looks exactly as the surface contribution in non-relativistic (or SRA) expression (\ref{sm_nr1}) if one interchanges the radial derivatives with small components. For applications, it is important to understand that in the last line of equation (\ref{sm_term}) the left component of each product comes from the original conjugated spinor. In order to clarify the statement, let us give couple of examples of evaluation of the $til\mu$-component of the surface term. In the first example, let both original spinors ($\Psi^{\dagger}$ and $\Psi$) belong to APW type, i.e. we are evaluating the surface correction to the matrix element of kinetic energy between 
$\Pi^{^{\dagger}\mathbf{k}}_{\mathbf{G}s}(\mathbf{r})$ and $\Pi^{\mathbf{k}}_{\mathbf{G}'s'}(\mathbf{r})$. In this case, the correspondence of the quantities appearing in the last line of (\ref{sm_term}) and realistic quantities is the following:

\begin{equation} \label{apw_apw}
\begin{array}{cc}
g^{^{*}t}_{il\mu}(S_{t}) \rightarrow y^{^{*}(1)\mathbf{k}}_{til \mu;\mathbf{G}s};&
f^{I}_{il\mu}(S_{t}) \rightarrow y^{(2)\mathbf{k}}_{til \mu;\mathbf{G}'s'}  \\
f^{^{*}t}_{il\mu}(S_{t}) \rightarrow f^{t}_{il}(S_{t})y^{^{*}(1)\mathbf{k}}_{til \mu;\mathbf{G}s};&
g^{I}_{il\mu}(S_{t}) \rightarrow y^{(1)\mathbf{k}}_{til \mu;\mathbf{G}'s'}  \\
g^{^{*}I}_{il\mu}(S_{t}) \rightarrow y^{^{*}(1)\mathbf{k}}_{til \mu;\mathbf{G}s};&
f^{t}_{il\mu}(S_{t}) \rightarrow f^{t}_{il}(S_{t})y^{(1)\mathbf{k}}_{til \mu;\mathbf{G}'s'}  \\
f^{^{*}I}_{il\mu}(S_{t}) \rightarrow y^{^{*}(2)\mathbf{k}}_{til \mu;\mathbf{G}s};&
g^{t}_{il\mu}(S_{t}) \rightarrow y^{(1)\mathbf{k}}_{til \mu;\mathbf{G}'s'}  \\
\end{array}.
\end{equation}

With the above correspondence, the total contribution is
\begin{eqnarray} \label{apw_apw_t}
-\frac{S^{2}_{t}}{2}&\Big[y^{^{*}(1)\mathbf{k}}_{til \mu;\mathbf{G}s} y^{(2)\mathbf{k}}_{til \mu;\mathbf{G}'s'}  -2f^{t}_{il}(S_{t})y^{^{*}(1)\mathbf{k}}_{til \mu;\mathbf{G}s}y^{(1)\mathbf{k}}_{til \mu;\mathbf{G}'s'}\nonumber\\&+y^{^{*}(2)\mathbf{k}}_{til \mu;\mathbf{G}s}y^{(1)\mathbf{k}}_{til \mu;\mathbf{G}'s'}  \Big],
\end{eqnarray}
which is used in deriving the equation (\ref{hmtaa2}) of the main text.

In the second example let us think that we are evaluating correction to the matrix element of local function of 'lo' type and $\Pi^{\mathbf{k}}_{\mathbf{G}'s'}(\mathbf{r})$ which now can be APW or LAPW type. In this case the correspondence is the following:

\begin{equation} \label{lo_apw}
\begin{array}{cc}
g^{^{*}t}_{il\mu}(S_{t}) \rightarrow 0;&
f^{I}_{il\mu}(S_{t}) \rightarrow y^{(2)\mathbf{k}}_{til \mu;\mathbf{G}'s'}  \\
f^{^{*}t}_{il\mu}(S_{t}) \rightarrow f^{(lo)t}_{il}(S_{t});&
g^{I}_{il\mu}(S_{t}) \rightarrow y^{(1)\mathbf{k}}_{til \mu;\mathbf{G}'s'}  \\
g^{^{*}I}_{il\mu}(S_{t}) \rightarrow 0;&
f^{t}_{il\mu}(S_{t}) \rightarrow f^{t}_{il}(S_{t})  \\
f^{^{*}I}_{il\mu}(S_{t}) \rightarrow 0;&
g^{t}_{il\mu}(S_{t}) \rightarrow y^{(1)\mathbf{k}}_{til \mu;\mathbf{G}'s'}  \\
\end{array}.
\end{equation}

With the above correspondence, the total contribution is
\begin{equation} \label{lo_apw_t}
\frac{S^{2}_{t}}{2}f^{(lo)t}_{il}(S_{t})y^{(1)\mathbf{k}}_{til \mu;\mathbf{G}'s'},
\end{equation}
which is used in deriving the equation (\ref{hmtba2}) of the main text.


\end{document}